\pdfoutput=1
\documentclass[12pt]{article}
\usepackage{cite}

.5 scaled \magstep4
\font\medio=cmr9.5 scaled \magstep2
\outer\def\beginsection#1\par{\medbreak\bigskip
      \message{#1}\leftline{\bf#1}\nobreak\medskip
\vskip-\parskip
      \noindent}
\usepackage{graphicx} 
\begin{document}
\bibliographystyle {unsrt}

\begin{center}
{\Large {\bf Inflation, space-borne interferometers}}\\
\vspace{3mm}
{\Large {\bf and the expansion history of the Universe}}\\
\vspace{15mm}
 Massimo Giovannini 
 \footnote{Electronic address: massimo.giovannini@cern.ch} \\
\vspace{1cm}
{{\sl Department of  Physics, CERN, 1211 Geneva 23, Switzerland }}\\
\vspace{0.5cm}
{{\sl INFN, Section of Milan-Bicocca, 20126 Milan, Italy}}

\vspace*{1cm}
\end{center}

\centerline{\medio  Abstract}
\vspace{5mm}
According to the common wisdom, between a fraction of the mHz and few Hz the spectral energy density of the inflationary gravitons can be safely disregarded even assuming the most optimistic sensitivities of the space-borne detectors. 
In this analysis we show that this conclusion is evaded if, prior to nucleosynthesis, the post-inflationary evolution includes a sequence of stages expanding either faster or slower than radiation. As a consequence, contrary to the conventional lore, it is shown that below a fraction of the Hz the spectral energy density of the relic gravitons may exceed (even by eight orders of magnitude) the signal obtained under the hypothesis of radiation dominance throughout the whole expansion history prior to the formation of light nuclei. Since the slopes and the amplitudes of the spectra specifically reflect both the inflationary dynamics and the subsequent decelerated evolution, it is possible to disentangle the contribution of the relic gravitons from other (late-time) bursts of gravitational radiation associated, for instance, with a putative strongly first-order phase transition at the TeV scale. Hence, any limit on the spectral energy density of the relic gravitons in the mHz range simultaneously constrains the post-inflationary expansion history and the inflationary initial data.
\vskip 0.5cm

\nonumber
\noindent

\vspace{5mm}

\vfill
\newpage

\renewcommand{\theequation}{1.\arabic{equation}}
\setcounter{equation}{0}
\section{Introduction}
\label{sec1}
A striking prediction of the early evolution of the space-time curvature is the formation of a stochastic background of relic gravitons  \cite{gr1,gr2,park} whose frequencies may extend between the aHz and the GHz regions. As originally pointed out in Ref. \cite{staro1}, the spectral energy density of relic gravitons is quasi-flat between $100$ aHz and $100$ MHz for the inflationary scenarios relying on the conventional slow-roll evolution. Since the quasi-flat plateau corresponds to wavelengths that left the Hubble radius during inflation and reentered after radiation was already dominant, the spectral energy density can only reach its maximum in the aHz region where the signal scales as $\nu^{-2}$ with the comoving frequency $\nu$ \cite{rub}. In the aHz interval the temperature and the polarization anisotropies of the cosmic microwave background are customarily 
employed to infer the tensor to scalar ratio $r_{T}$ here assumed in the range $r_{T}= r_{T}(\nu_{p})\leq 0.06$, as suggested by recent determinations \cite{TS1,TS2,TS3}. For the record $\nu_{p} = k_{p}/(2\pi) = 3.09\,\mathrm{aHz}$ and  $k_{p} = 0.002 \,\, \mathrm{Mpc}^{-1}$ denotes the pivot scale at which the scalar and tensor power spectra are conventionally assigned when the relevant wavelengths are larger than the Hubble radius prior to matter-radiation equality.

Depending on the value of $r_{T}$, the spectral energy density in critical units\footnote{Instead of working with the spectral 
energy density of the relic gravitons in critical units (conventionally denoted by $\Omega_{gw}(\nu,\tau_{0})$) it is practical to introduce $h_{0}^2\, \Omega_{gw}(\nu,\tau_{0})$ where $h_{0}$ is the indetermination of the Hubble rate. The spectral energy density of the relic gravitons {\em does not} 
coincide with their energy density in critical units which is instead frequency-independent. We also note that the frequencies are often mentioned in the text by using the standard metric prefixes of the 
international system of units. So, for instance, $\mathrm{aHz} = 10^{-18}\,\, \mathrm{Hz}$, $\mathrm{mHz} = 10^{-3} \,\, \mathrm{Hz}$ and so on. After the analysis 
of Ref. \cite{staro1} suggesting a flat slope for $\Omega_{gw}(\nu,\tau_{0})$ various authors discussed the same problem with a number of relevant additions; in this respect the interested 
reader may consult Refs. \cite{ADD1a,ADD1b,ADD2,ADD3,ADD4}.} well above $100$ aHz is $h_{0}^2 \Omega_{gw}(\nu, \tau_{0}) < {\mathcal O}(10^{-17})$ and this estimate includes the effect of the various damping source such as 
the late-dominance of the dark energy, the evolution of the relativistic species 
and the free-streaming of the neutrinos \cite{STRESSNU1,STRESSNU2}.
For all these reasons the spectral energy density of inflationary origin is too 
small to be detected by either ground based or space-borne 
detectors even in their most advanced versions. At the moment the only direct bounds on the relic gravitons come from the audio band and depend upon the spectrum of the signal but we could anyway say that, for a nearly scale-invariant spectrum,  $h_{0}^2\Omega_{gw}(\nu, \tau_{0}) < {\mathcal O}(10^{-9})$  between $10$ Hz and $80$ Hz \cite{LIGO1,LIGO2} (see also \cite{LIGO3} for a recent review including earlier bounds). 
For a physical comparison between the  ground-based detectors and the (futuristic) space-borne interferometers the spectral energy density can be usefully expressed in terms of the chirp amplitude $h_{c}(\nu,\tau_{0})$ \cite{LIGO3} when the typical frequencies fall in the audio band:
 \begin{equation}
h_{0}^2 \Omega_{gw}(\nu,\tau_{0}) = 6. 26 \times 10^{-9} \biggl(\frac{\nu}{0.1 \, \, \mathrm{kHz}}\biggr)^2 
\biggl[\frac{h_{c}(\nu,\tau_{0})}{10^{-24}} \biggr]^2.
\label{ONEeq}
\end{equation}
If we read Eq. (\ref{ONEeq}) from left to right  we can argue that to probe
$h_{0}^2 \Omega_{gw}(\nu,\tau_{0}) = {\mathcal O}(10^{-9})$ we would 
need a sensitivity in the chirp amplitude ${\mathcal O}(10^{-24})$ for a typical frequency $\nu = {\mathcal O}(100)$ Hz.  From right to left Eq. (\ref{ONEeq}) suggests instead that, for the same sensitivity 
in $h_{c}(\nu,\tau_{0})$, the minimal detectable $h_{0}^2 \Omega_{gw}(\nu,\tau_{0})$ gets 
comparatively smaller\footnote{Besides the the absence of seismic noise this is probably one of strongest arguments  in favour of space-borne detectors 
for typical frequencies ranging between a fraction of the mHz and the Hz. }. This is why the 
minimal detectable spectral energy density could be $h_{0}^2 \Omega_{gw}(\nu,\tau_{0}) = {\mathcal O}(10^{-11})$ or even $h_{0}^2 \Omega_{gw}(\nu, \tau_{0}) = {\mathcal O}(10^{-15})$ under the hypothesis that the same sensitivity reached in the audio band for the chirp amplitude can also be achieved in the mHz range. With this hope, various space-borne 
detectors have been proposed so far: the Laser Interferometric Space Antenna (LISA) \cite{LISA1,LISA2}, 
 the Deci-Hertz Interferometer Gravitational Wave Observatory (DECIGO) \cite{DECIGO1,DECIGO2},  the Ultimate-DECIGO \cite{UDECIGO} (conventionally 
 referred to as U-DECIGO), the Big Bang Observer (BBO) \cite{BBO}. This list has been recently enriched by the Taiji \cite{TAIJI1,TAIJI2} and by the TianQin \cite{TIANQIN1,TIANQIN2} experiments. Since these instruments are not yet operational (but might come into operation within the next twenty years) their actual sensitivities are difficult to assess, at the moment. However, without dwelling on the specific nature 
 of the noise power spectra, Eq. (\ref{ONEeq}) shows that, as long as $h_{c} = {\mathcal O}(10^{-23})$ the space-borne detectors might probe $h_{0}^2 \Omega_{gw}(\nu, \tau_{0}) = {\mathcal O}(10^{-14})$ for $\nu_{S} = {\mathcal O}(0.01)$ Hz and this is, roughly 
 speaking, the daring expectation of DECIGO \cite{DECIGO1,DECIGO2} and of U-DECIGO \cite{UDECIGO}.

According to the standard lore (see e.g. \cite{LISA1,LISA2,DECIGO1,DECIGO2}) the astrophysical sources of gravitational radiation (i.e. mostly white dwarves and solar masses black holes) dominate the signal below $0.1$ Hz, while the bursts of gravitons from the TeV physics are unlikely in the standard electroweak theory but should be anyway subleading in comparison with the galactic foregrounds. Because of the relative smallness of its spectral energy density, the inflationary background of relic gravitons is always disregarded but this conclusion is only based on a specific expansion history and it can be evaded if, prior to nucleosynthesis, the evolution of the background is not constantly dominated by radiation. Indeed, the flatness of $h_{0}^2 \Omega_{gw}(\nu,\tau_{0})$ for frequencies larger than $100$ aHz is not only determined by 
the inflationary evolution when the relevant wavelengths exit the Hubble radius 
but also by the expansion rate at reentry \cite{ST1,ST2}.  
The high-frequency signal is maximized by   
a long stage expanding at a rate that is slower than radiation \cite{ST1,ST2} and this possibility 
is realized in various classes of quintessential inflationary scenarios \cite{PV,LL1,LL2,LL3} (see also \cite{ford,spok}). The signal from a long stiff phase does not imply a reduction of $h_{0}^2 \Omega_{gw}(\nu,\tau_{0})$ in the aHz region so that the high-frequency measurements of wide-band detectors and the low-frequency determinations of $r_{T}$ can be 
simultaneously constrained within an accurate numerical framework \cite{ST3,ST3a}. In this context the potential signal might be sufficiently large both in the aHz region and in the audio band. 

While in the case of a stiff post-inflationary phase the spike typically arises 
for frequencies between the GHz and $100$ GHz it is also possible to have 
different profiles of the spectral energy density with a number of 
different peaks when the frequency is comparatively smaller or even much smaller 
than the MHz. There is then a trade-off between the smallness 
of the frequency and the magnitude of $h_{0}^2 \Omega_{gw}(\nu,\tau_{0})$ \cite{ST3,ST3a}. Since the most general post-inflationary expansion rate consists of a series of successive stages expanding at different rates that are either faster 
or smaller than radiation\footnote{ Incidentally, within the present approach 
the possibility of a signal in the nHz band (recently suggested by the pulsar timing arrays \cite{CCPP1,CCPP2,CCPP3,NANO1}) has been specifically scrutinized by considering a wide range of possibilities including the presence of late-time stages of inflationary expansion \cite{ST4}. In this paper we are instead concerned with the mHz range and the potential signal from pulsar timing arrays will not be specifically discussed.} \cite{ST3,ST3a,ST4}, in this paper we are going to argue that the general approach previously 
explored is also applicable also to smaller frequencies in the mHz region. In the presence 
of a modified post-inflationary expansion rate  the standard inflationary signal computed in Refs. \cite{staro1,rub} can be much larger below the Hz and potentially dominant against the bursts of gravitational radiation from strongly first-order phase transitions.

The layout of this paper is the following. In section \ref{sec2}
 the inflationary power spectra are computed after the relevant 
wavelengths reentered the Hubble radius
during a post-inflationary stage that differs from radiation. 
In section \ref{sec3} we examine the general case where each stage 
of a  larger sequence of phases expands at a rate that is either faster or slower than radiation. 
In this situation the spectral energy density exhibits a succession of peaks and throughs 
whose frequencies are solely determined by the curvature scale.
Since the slopes of of the humps in $h_{0}^2 \Omega_{gw}(\nu,\tau_{0})$ depend both on the inflationary stage and on the post-inflationary evolution, in section \ref{sec4} it is shown that the current limits from ground-based detectors already pin down a well defined region of the parameter space that should be further explored by space-borne interferometers. Section \ref{sec5} contains the concluding remarks and some comments on the future perspectives.

\renewcommand{\theequation}{2.\arabic{equation}}
\setcounter{equation}{0}
\section{Spectral energy density of the inflationary gravitons}
\label{sec2}
The effect of the post-inflationary evolution is not, as sometimes 
argued, a purely kinematical problem that is virtually
disentangled from the dynamical evolution of the tensor modes. On the contrary the enhancement of the spectral energy density at late 
times is also determined by the early  expansion: it is 
because of the successive occurrence of an inflationary stage and of the late post-inflationary evolution that $\Omega_{gw}(k,\tau)$ may be enhanced at high and intermediate frequencies \cite{ST1,ST2}.  The flat spectrum of relic gravitons for frequencies larger than
$100$ aHz only arises if the relevant wavelengths exit the Hubble radius 
during inflation and reenter in a radiation-dominated stage of expansion, as 
originally assumed in Refs. \cite{staro1,rub}. We may consider, in this respect the standard form of the spectral energy density in critical units that can be written as \cite{LIGO3}:
\begin{equation}
\Omega_{gw}(k,\tau_{0}) = \frac{1}{24 H^2 \, a^2} \biggl[ Q_{T}(k,\tau) + k^2 \, P_{T}(k,\tau) \biggr],
\label{SPendens1}
\end{equation}
where $a(\tau)$ is the scale factor of a conformally flat background geometry, 
$\tau$ is the conformal time and $H$ is the standard Hubble expansion rate. In Eq. (\ref{SPendens1}) $Q_{T}(k,\tau)$ and $P_{T}(k,\tau)$ are the tensor power spectra 
that are defined from the evolution of the mode functions $G_{k}(\tau)$ and $F_{k}(\tau)$: 
\begin{equation}
Q_{T}(k,\tau) = \frac{ 4 \ell_{P}^2}{ \pi^2} \, k^3\, \bigl|\, G_{k}(\tau)\bigr|^2,\qquad\qquad P_{T}(k,\tau) = \frac{ 4 \ell_{P}^2}{ \pi^2} \, k^3\, \bigl|\, F_{k}(\tau)\bigr|^2,
\label{SP2}
\end{equation}
where $\ell_{P} = \sqrt{ 8 \, \pi \, G}$; in what follows the notations 
for the Planck mass are given by $\overline{M}_{P} = M_{P}/\sqrt{8 \, \pi} = 1/\ell_{P}$
and $\overline{M}_{P}$ is the reduced Planck mass.

The rescaled mode functions 
 $f_{k}(\tau) =a(\tau) F_{k}(\tau)$ and $g_{k}(\tau) = a(\tau) G_{k}(\tau)$
 obey, in the present context, the standard evolution equations:
 \begin{equation}
 f_{k}^{\prime\prime} + \biggl[ k^2 - \frac{a^{\prime\prime}}{a} \biggr] f_{k} =0, \qquad \qquad g_{k} = f_{k}^{\prime} - {\mathcal H} f_{k}.
 \label{MFeq}
 \end{equation}
In Eq. (\ref{MFeq}) the prime denotes a derivation with respect to the conformal time coordinate $\tau$; we also use the standard notation  ${\mathcal H} = a H$ 
 where ${\mathcal H} = a^{\prime}/a$ and $H$ is the conventional Hubble 
 rate. Within the WKB approximation Eq. (\ref{MFeq}) is  
 approximately solved in the two complementary regimes where $k^2$   
is  either larger or smaller than $|\, a^{\prime\prime}/a|$. In particular 
when $k^2 \gg |\, a^{\prime\prime}/a\,|$ the mode functions 
($f_{k}$, $g_{k}$) oscillate while ($F_{k}$, $G_{k}$) are also 
suppressed as $1/a$. In the opposite regime (i.e. $k^2 \ll |\, a^{\prime\prime}/a|$) 
$f_{k}(\tau)$ is said to be superadiabatically amplified, according 
to the terminology originally introduced in Refs. \cite{gr1,gr2}. 
The oscillating and the superadiabatic regimes are separated by a 
region where the solutions change their analytic behaviour and these 
turning points are defined as solutions of the approximate equation $k^2 \simeq 
|\,a^{\prime\prime}/a|$ that can also be rewritten as: 
\begin{equation}
k^2 = a^2 \, H^2 \biggl[ 2 - \epsilon(a)\biggr], \qquad \qquad \epsilon(a) = - \frac{\dot{H}}{H^2}.
\label{turn1}
\end{equation}
During the inflationary stage of expansion $\epsilon\ll 1$ denotes the standard 
slow-roll parameter; conversely in the post-inflationary phase 
the background decelerates (but still expands) and $\epsilon(a) = {\mathcal O}(1)$.
If $\epsilon \neq 2$ both turning points are regular and this means that 
Eq. (\ref{turn1}) can be approximately solved by $k \simeq a H $. For 
instance when a given wavelength crosses the Hubble radius during 
inflation we have that $\epsilon \ll 1$ and $k \simeq a_{ex} \, H_{ex}$ 
that also means, by definition, $k \tau_{ex} \simeq 1$. Similarly 
if the given wavelength reenters in a decelerated stage of expansion different from radiation we also have that $k \simeq a_{re} \, H_{re}$. Finally if the reentry occurs in the radiation stage we have that $\epsilon_{re}\to 2$ and the condition (\ref{turn1}) implies that $k \tau_{re} \ll 1$. 

These considerations suggest that the spectral energy density of the relic gravitons 
depends both on the exit and on the reentry of the given wavelength and for this purpose it is appropriate to express the mode functions in the Wentzel-Kramers-Brillouin (WKB) approximation under the further assumption that $a_{re} \gg a_{ex}$: this requirement is
verified as long as the Universe expands as it is aways the case throughout the
present discussion. The initial conditions for the evolution of the mode 
functions are then assigned during the inflationary stage and before 
the corresponding wavelengths exit the Hubble radius; in this regime 
$f_{k}(\tau)$ and $g_{k}(\tau)$ are simply plane waves obeying the Wronskian 
normalization condition:
\begin{equation}
f_{k}(\tau) \, g_{k}^{\ast}(\tau) - f_{k}^{\ast}(\tau) g_{k}(\tau) = i,
\end{equation}
as required by the canonical commutation relations of the corresponding field operators \cite{LIGO3}. From the continuity of the mode functions 
across the turning points of the problem, the expression of $F_{k}(\tau)$ becomes:
\begin{equation}
F_{k}(\tau) = \frac{e^{- i k \, \tau_{ex}}}{a \sqrt{2 \, k}} \, {\mathcal Q}_{k}(\tau_{ex}, \tau_{re}) \, \biggl(\frac{a_{re}}{a_{ex}}\biggr) \biggl\{ \frac{{\mathcal H}_{re}}{k} \sin{[k ( \tau- \tau_{re})]}
+ \cos[k (\tau - \tau_{re})] \biggr\},
\label{MFW1}
\end{equation}
and it is valid for $k \tau \gg 1$ when all the corresponding 
wavelengths are shorter than the Hubble radius. In the same 
approximation $G_{k}(\tau)$ is:
\begin{equation}
G_{k}(\tau) = \frac{e^{- i k \, \tau_{ex}}}{a}\,\, \sqrt{\frac{k}{2}}\,\, {\mathcal Q}_{k}(\tau_{ex}, \tau_{re}) \, \biggl(\frac{a_{re}}{a_{ex}}\biggr) \biggl\{ \frac{{\mathcal H}_{re}}{k} \cos{[k ( \tau- \tau_{re})]}
- \sin[k (\tau - \tau_{re})] \biggr\}.
\label{MFW2}
\end{equation}
Equations (\ref{MFW1})--(\ref{MFW2}) are valid in the two concurrent limits $k\tau \gg 1$ and $a_{re} \gg a_{ex}$ but they are otherwise general since the expansion rates at $\tau_{ex}$ and $\tau_{re}$ have not been specified. 
In Eqs. (\ref{MFW1})--(\ref{MFW2}) ${\mathcal Q}_{k}(\tau_{ex}, \tau_{re})$ denotes 
a complex amplitude defined as: 
\begin{equation}
{\mathcal Q}_{k}(\tau_{re}, \tau_{ex}) = 1 - ( i \, k + {\mathcal H}_{ex}) \int_{\tau_{ex}}^{\tau_{re}} \frac{a_{ex}^2}{a^2(\tau)} \,\, d\, \tau.
\label{MFW3}
\end{equation}
The integral at the right hand side of Eq. (\ref{MFW3}) depends on the evolution 
between $\tau_{ex}$ and $\tau_{re}$ but its value is always subleading 
so that it is generally true\footnote{Only if $a^2(\tau) \simeq 1/{\mathcal H}$ the contribution of the integral of Eq. (\ref{MFW3}) is relevant and it corresponds to the possibility of extended stiff phases where, for instance, the energy density is dominated by the kinetic energy of a scalar field \cite{ST1,ST2}; in this case the spectral energy density and the other observables inherit a logarithmic correction.} that $\bigl|{\mathcal Q}_{k}(\tau_{re}, \tau_{ex})\bigr|^2 \to 1$. 

In Eqs. (\ref{MFW1})--(\ref{MFW2}) we may note the appearance of standing waves that are characteristic both in the case of relic gravitons 
and in the case of scalar metric fluctuations and they are often referred to as Sakharov oscillations because they arose, for the first time, in the pioneering contribution of Ref. \cite{SAK1} (see also \cite{SAK2}). When Eqs. (\ref{MFW1})--(\ref{MFW2}) are inserted into Eq. (\ref{SP2}) we can obtain the corresponding power spectra that determine the final expression of the spectral energy density in critical units through Eq. (\ref{SPendens1}):
\begin{equation}
\Omega_{gw}(k,\tau) = \frac{k^4}{12\, \pi^2 \,a^4\, H^2 \, \overline{M}_{P}^2} \bigl| {\mathcal Q}(\tau_{ex}, \tau_{re})\bigr|^2 \biggl(\frac{a_{re}}{a_{ex}}\biggr)^2 \biggl( 1 + \frac{{\mathcal H}_{re}^2}{k^2}\biggr) \biggl[ 1 + {\mathcal O}\biggl(\frac{{\mathcal H}}{k}\biggr)\biggr].
\label{SPen2}
\end{equation}
Equation (\ref{SPen2}) is valid in the limit ${\mathcal H}/k \ll 1$ and this  condition is 
equivalent to $k \tau \gg 1$ since ${\mathcal H} ={\mathcal O}(\tau^{-1})$. If a given wavelength exits the Hubble radius during inflation we have:
\begin{equation}
k \simeq a_{ex} \, H_{ex} \simeq - \frac{1}{( 1 - \epsilon) \tau_{ex}} = - \frac{\beta}{\tau_{ex}},
\label{SPen2a}
\end{equation}
where we denoted, for the sake of convenience, $\beta = 1/(1 - \epsilon)$ and $\epsilon_{ex} = \epsilon \ll 1$. When the same wavelength reenters  
during a stage that is {\em not} dominated by radiation, $\epsilon_{re} \neq 2$ in Eq. (\ref{turn1}) so that, at reentry,  $k \simeq {\mathcal H}_{re} = a_{re} \, H_{re}$.

If a given wavelength $2\pi/k$ 
reenters across two different regimes characterized 
by a different expansion rate, the scale factor during the $i$-th stage of expansion 
can be parametrized, for instance,  as: 
\begin{equation}
a_{i}(\tau) =  \biggl(\frac{\tau}{\tau_{i}}\biggr)^{\delta_{i}}, \qquad \delta_{i} >0, \qquad \delta_{i} \neq 1, \qquad \tau \leq \tau_{i}.
\label{ai}
\end{equation}
For $\tau > \tau_{i}$ the scale factor during the 
$(i+1)$-th stage of expansion is: 
modified as 
\begin{equation}
a_{i+1}(\tau) \simeq \biggl[\frac{\delta_{i}}{\delta_{i+1}} \biggl(\frac{\tau}{\tau_{i}}-1\biggr)+1 \biggr]^{\delta_{i+1}}, \qquad \delta_{i+1} >0, \qquad \delta_{i+1} \neq 1, \qquad \tau \geq \tau_{i}.
\label{aipl1}
\end{equation}
Even if $\delta_{i+1}$ and $\delta_{i}$ can be equal, the situation we want to discuss
now is the one where $\delta_{i+1} \neq \delta_{i}$. Let us then go back to Eq. (\ref{SPen2}) and evaluate the spectral energy density for the modes reentering for $\tau< \tau_{i}$ 
\begin{equation}
\Omega^{(i)}_{gw}(k,\tau) = \frac{4}{3 \pi} \biggl(\frac{H_{1}}{M_{P}}\biggr)^2 \, \biggl( \frac{H_{1} \, a_{1}^2}{ H\, a^2}\biggr)^2 \biggl|\frac{a_{i} H_{i}}{a_{1}\, H_{1}} \biggr|^{4 - \beta} \, \biggl| 
\frac{k}{a_{i} \, H_{i}} \biggr|^{n_{T}^{(i)}},\qquad\qquad k > a_{i}\, H_{i},
\label{OMgwi}
\end{equation}
where we took into account that $k \simeq a_{re} H_{re} = {\mathcal H}_{re}$. 
In Eq. (\ref{OMgwi}) $H_{1} $ is coincides with the maximal value of the Hubble rate 
(e.g. at the end of inflation). We stress that, in the case $\epsilon_{re} \neq 2$, the turning points of Eq. (\ref{turn1}) are determined from $k^2 \simeq a^2\, H^2$; the contribution 
of the numerical factor $(2 -\epsilon_{ex})$ and $(2 - \epsilon_{re})$ 
has been consistently neglected.  The spectral index appearing in Eq. (\ref{OMgwi})
is given by 
\begin{equation}
n_{T}^{(i)} = 2 ( 1 - \beta) + 2 ( 1 - \delta_{i}), \qquad \delta_{i} > 0, \qquad \beta= \frac{1}{1 -\epsilon}.
\end{equation} 
If we now assume the validity of the consistency relations the value of $n_{T}^{(i)}$ depends 
on $\delta_{i}$ and $r_{T}$
 \begin{equation} 
n_{T}^{(i)}(r_{T}, \delta_{i}) = \frac{32 - 4 \, r_{T}}{16 -r_{T}} - 2 \delta_{i} = 
2( 1 - \delta_{i}) + {\mathcal O}(r_{T}).
\label{THREE2}
\end{equation}
From Eq. (\ref{THREE2}) in the limit $\delta_{i}\to 1$ we have 
\begin{equation}
\lim_{\delta_{i} \to 1} \, \, n_{T}^{(i)}(r_{T},\delta_{i}) = - r_{T}/8 + {\mathcal O}(r_{T}^2).
\label{THREE2aa}
\end{equation}
Similarly, for the wavelengths reentering during the $(i+1)$-th stage the spectral energy density is instead given by:
\begin{equation}
 \Omega^{(i+1)}_{gw}(k,\tau) = \frac{4}{3 \pi} \biggl(\frac{H_{1}}{M_{P}}\biggr)^2 \, \biggl( \frac{H_{1} \, a_{1}^2}{ H\, a^2}\biggr)^2 \biggl|\frac{a_{i} H_{i}}{a_{1}\, H_{1}} \biggr|^{4 - \beta} \, \biggl| 
\frac{k}{a_{i} \, H_{i}} \biggr|^{n_{T}^{(i+1)}},\qquad\qquad k < a_{i}\, H_{i},
\label{OMgwiPLUS}
\end{equation}
where $n_{T}^{(i+1)}$ is now given by
\begin{equation}
 n_{T}^{(i+1)}(r_{T}, \delta_{i+1}) = 2 ( 1 - \beta) + 2 ( 1 - \delta_{i+1}) =
 \frac{32 - 4 \, r_{T}}{16 -r_{T}} - 2 \delta_{i+1} = 
2( 1 - \delta_{i+1}) + {\mathcal O}(r_{T}).
\label{THREE2a}
 \end{equation}
  From Eqs. (\ref{THREE2})--(\ref{THREE2a}) we have in fact three complementary possibilities. If the expansion rate is initially slower than radiation (i.e. $\delta_{i} < 1$)
the spectral index for $k > a_{i}\, H_{i}$ is either blue or violet (i.e. $n_{T}^{(i)} > 0$). 
Then, provided $\delta_{i + 1} > 1$ the spectral energy density of Eqs. (\ref{OMgwi}) and (\ref{OMgwiPLUS}) is characterized by a local minimum for $k \simeq a_{i}\, H_{i}$. 
We may also have the opposite situation where the expansion rate is initially 
faster than radiation (i.e. $\delta_{i} >1$) then it gets smaller (i.e. $\delta_{i+1} <1$)
and the spectral energy density has a local maximum always for $k \simeq a_{i} \, H_{i}$.
The third possibility suggests that either $\delta_{i} \to 1$ or $\delta_{i+1} \to 1$: in this 
case the spectral energy density exhibits a quasi-flat branch either 
for $\nu < \nu_{i} = a_{i} \, H_{i}$ or for $ \nu> \nu_{i}$.
While this result is formally correct, for the sake of completeness it is 
useful to verify it directly from Eq. (\ref{SPen2}). 
In the case $\delta_{i} \to 1$ at reentry we have  that $k \tau_{re} \ll 1$; this means 
that the term ${\mathcal H}_{re}/k \gg 1$ dominates in Eq. (\ref{SPen2}) and  $\Omega_{gw}(k,\tau)$ 
is then given by:
\begin{equation}
\Omega_{gw}(k,\tau) = \frac{2}{3\pi} \biggl(\frac{k^2}{a_{ex}^2 M_{P}^2}\biggr) \biggl(\frac{H_{re}^2 \, a_{re}^4}{H^2 \, a^4}\biggr) =  \frac{2}{3\pi} \biggl(\frac{H_{1}^2 \, a_{1}^4}{H^2 \, a^4}\biggr) \biggl(\frac{H_{1}}{M_{P}}\biggr)^2 \, \biggl(\frac{k}{a_{1}\, H_{1}}\biggr)^{\overline{n}_{T}},
\label{conf}
\end{equation}
where $\overline{n}_{T} = - 2 \epsilon \simeq - r_{T}/8$. Thus Eqs. (\ref{THREE2aa}) and (\ref{conf}) show that $n_{T}^{(i)}(r_{T}, \delta_{i})$ evaluated in the limit $\delta_{i} \to 1$ 
indeed corresponds to $\overline{n}_{T}$ up to corrections ${\mathcal O}(r_{T}^2)$.

\renewcommand{\theequation}{3.\arabic{equation}}
\setcounter{equation}{0}
\section{Peaks and throughs of the spectral energy density}
\label{sec3}
The previous results demonstrate
that ${\mathcal H}_{re}$ and ${\mathcal H}_{ex}$  
are equally essential for the late-time form of the spectral energy density. In this sense the slopes of the humps appearing in $\Omega_{gw}(\nu,\tau_{0})$
are a simultaneous test of the expansion rate during inflation 
and in the post-inflationary stage. This means that if $n_{T}^{(i)}(r_{T}, \delta_{i})$ and $n_{T}^{(i+1)}(r_{T}, \delta_{i+1})$ are observationally assessed around a given peak, their measurement
ultimately reflects the expansion history during and after inflation.

\subsection{The profile of the effective expansion rate}

\begin{figure}[!ht]
\centering
\includegraphics[height=9cm]{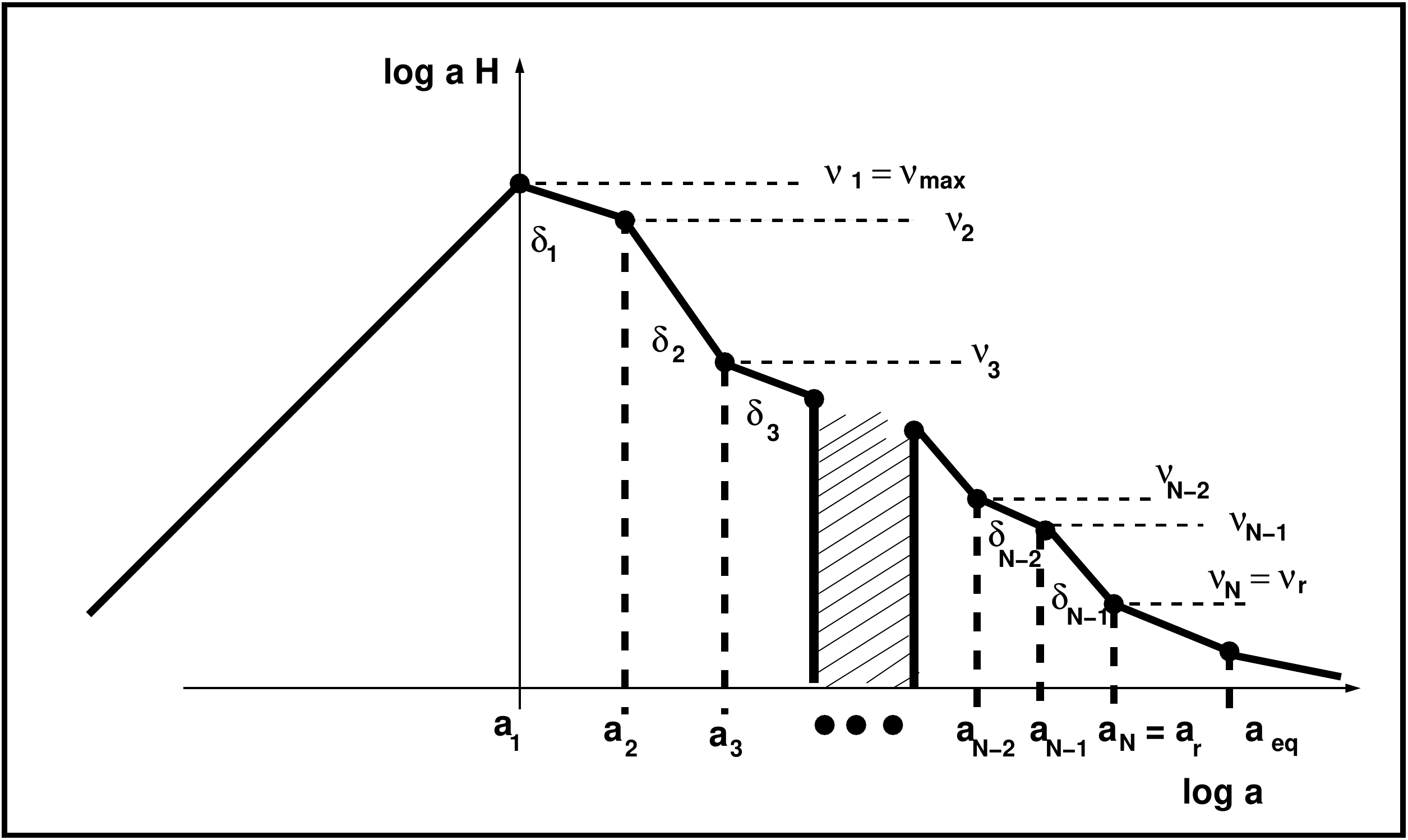}
\caption[a]{On the vertical axis the common logarithm of $a\, H$ is illustrated as a function 
of the common logarithm of the scale factor. We consider here the general situation where 
there are $N$ different stages of expansion not necessarily coinciding 
with radiation. The $N$-th stage conventionally coincides with 
the standard radiation-dominated evolution (i.e. $a_{r} = a_{N}$) while 
the first stage starts at the end of inflation (i.e. $H_{1} = H_{max}$). Even if, in general, $H_{r} \ll H_{1}$ 
we shall always require $H_{r} >10^{-44} \, M_{P}$ implying that the dominance 
of radiation takes place well  before big-bang nucleosynthesis. During the $i$-th stage 
of the sequence the scale factor expands as $a(\tau) \simeq \tau^{\delta_{i}}$. The dashed 
lines appearing in this cartoon correspond to the pivotal frequencies of the spectrum.}
\label{FIG1}      
\end{figure}

It is now interesting to consider the general case illustrated in Fig. \ref{FIG1} 
where, prior to $a_{1}$, an inflationary 
stage  dominates the evolution of the background 
so that the effective expansion rate $a\, H$ increases 
linearly with the scale factor.  As suggested in the previous section, during this stage the initial inhomogeneities of the tensor modes are normalized to their quantum 
mechanical values. While in the conventional case we would have that, after inflation, the effective expansion rate is immediately dominated by radiation, 
in the situation illustrated in Fig. \ref{FIG1} we rather consider 
a sequence of different stages expanding 
either faster or slower than radiation.
More specifically, according to Fig. \ref{FIG1}, the post-inflationary expansion history consists of $N$ successive stages where, by definition, $a_{1}$ coincides with the end of inflation. Moreover, since $a_{r}$ denotes the value of the scale factor at the onset of the radiation-dominated stage of expansion, we conventionally posit that 
$a_{N}= a_{r}$. During each of the successive stages the expansion rate is characterized, in the conformal time 
coordinate, by $a(\tau) \sim \tau^{\delta_{i}}$  so that the spectral index 
of $h_{0}^2 \Omega_{gw}(\nu,\tau_{0})$ 
is in fact the one already determined in Eqs. (\ref{THREE2}) and (\ref{THREE2a}). 
During the $i$-th stage of expansion the spectral energy density in critical units 
 scales approximately as $(\nu/\nu_{i})^{n_{T}^{(i)}}$.
The dashed lines of Fig. \ref{FIG1} illustrate the 
values of the comoving frequencies at the transition points. Since  the current
value of the scale factor is conventionally normalized to $1$ (i.e. $a_{0} =1$) comoving and physical frequencies 
coincide at the present time but not earlier on. Furthermore the largest frequency 
 coincides with $a_{1} \,H_{1}$ while for 
$\nu< \nu_{r} = a_{r} H_{r}$ the spectral energy density has the standard 
quasi-flat form since the corresponding wavelengths exit the Hubble radius 
during inflation and reenter when the Universe is already dominated by radiation.
It is important to appreciate that while $\nu_{1} = a_{1} \, H_{1} = \nu_{max}$ 
depends on all the post-inflationary expansion rates (i.e. the different $\delta_{i}$),
$\nu_{r} = a_{r} \, H_{r}$ only depends on the hierarchy between $H_{1}$ and $H_{r}$. 
To prove this statement it is practical to introduce the ratios of the curvature scales during two successive 
stages of expansion, namely:
\begin{equation}
\xi_{i} = \frac{H_{i+1}}{H_{i}} < 1, \qquad \xi = \prod_{i}^{N-1} \xi_{i} =  \frac{H_{r}}{H_{1}} <1,
\label{THREE3}
\end{equation}
where $\xi$ denotes the ratio between the Hubble rates at the onset 
of the radiation-dominated stage\footnote{A curvature scale $H_{r} = {\mathcal O}(10^{-44})\, M_{P}$ correspond to a temperature of the plasma $T= {\mathcal O}(\mathrm{MeV})$. For the present 
ends it is more practical to work directly with the curvature scales.}  (i.e. $H_{r}$) and at the end of the inflationary phase (i.e. $H_{1}$);  $\xi_{i}$ gives instead the ratio of the expansion rates between two 
successive stages. Note that, by definition, both $\xi_{i}$ and $\xi$ are smaller than $1$ since the largest value of the Hubble rate always appears in the 
denominator. Since  we conventionally choose that $H_{1}$ concides with the expansion rate 
at the end of inflation (i.e. $H_{1} \equiv H_{max}$) while $H_{N} \equiv H_{r}$,
in the simplest non-trivial situation we have that $N=3$ and Eq. (\ref{THREE3}) implies:
\begin{equation}
\xi = \xi_{1} \, \xi_{2} = \frac{H_{r}}{H_{1}}, \qquad \xi_{1} = \frac{H_{2}}{H_{1}}, \qquad 
\xi_{2} = \frac{H_{r}}{H_{2}},
\end{equation}
where, following the conventions established established above and illustrated in Fig. \ref{FIG1}, $a_{3}= a_{r}$ and $H_{3} = H_{r}$. 

\subsection{The typical frequencies of the spectrum}
As anticipated the maximal frequency of the spectrum indeed depends upon all the 
successive stages of expansion and its the general expression is:
\begin{equation}
\nu_{1} = \nu_{max} =  \prod_{i =1}^{N-1} \, \, \xi_{i}^{\frac{\delta_{i} -1}{2(\delta_{i} +1)}}\, \, \overline{\nu}_{max}.
\label{nuone}
\end{equation}
In Eq. (\ref{nuone}) $\overline{\nu}_{max}$ denotes the maximal frequency of the spectrum 
when all the different stages of expansion appearing in Fig. \ref{FIG1} 
collapse to a single phase expanding exactly like radiation. Indeed, if $\delta_{i} \to 1$ 
in Eq. (\ref{nuone}) for all the $i= 1, \, .\, ., \,. \, N$ we have that $\nu_{1} = \nu_{max} \to  \overline{\nu}_{max}$:
\begin{equation}
\overline{\nu}_{max}= 269.33 \,\biggl(\frac{r_{T}}{0.06}\biggr)^{1/4} \,\biggl(\frac{{\mathcal A}_{{\mathcal R}}}{2.41\times 10^{-9}}\biggr)^{1/4} \, \biggl(\frac{h_{0}^2 \, \Omega_{R0}}{4.15 \times 10^{-5}}\biggr)^{1/4} \,\,\,\mathrm{MHz},
\label{numax}
\end{equation}
where $\Omega_{R0}$ is the present fraction of relativistic species of the concordance scenario 
and ${\mathcal A}_{{\mathcal R}}$ is the amplitude of the scalar power spectrum that determines\footnote{We recall 
that $(H_{1}/M_{P}) = \sqrt{\pi \, \epsilon\, {\mathcal A}_{{\mathcal R}}}$. If the consistency relations 
are enforced we also have that $(H_{1}/M_{P}) = \sqrt{\pi \, r_{T}\, {\mathcal A}_{{\mathcal R}}}/4$, where $r_{T} \simeq 16 \, \epsilon$ 
is, as usual, the tensor to scalar ratio.}
$H_{1}$.
In other words $\overline{\nu}_{max}$ coincides with the maximal frequency 
of the spectrum in the case considered in Ref. \cite{staro1} where 
$H_{1} \to H_{r}$ and $ \nu_{r} \to \nu_{max} = {\mathcal O}(200)$ MHz.
In the general case illustrated in Fig. \ref{FIG1} 
we have that:
 \begin{equation}
 \nu_{N} = \nu_{r} = \prod_{j = 1 }^{N-1} \, \sqrt{\xi_{j}} \,\,\, \overline{\nu}_{max} = \sqrt{\xi} \, \overline{\nu}_{max},
 \label{nur}
 \end{equation}
where the second equality follows since, by definition, 
\begin{equation}
\prod_{j = 1 }^{N-1} \, \xi_{j}= \xi_{1}\,\xi_{2} \,.\,.\,. \,\xi_{N-2}\, \xi_{N-1} = \xi.
\label{nur2}
\end{equation}
  Equations (\ref{nuone}) and (\ref{nur}) demonstrate, 
as anticipated above, that while $\nu_{max}$ is sensitive to the 
whole expansion history, $\nu_{r}$ only depends upon $\sqrt{\xi}$ (where  $\xi= H_{r}/H_{1}$).

For all the other intermediate frequencies between $\nu_{max}$ and $\nu_{r}$,
the following expression holds: 
\begin{equation}
\nu_{m} = \prod_{j=1}^{m-1} \sqrt{\xi_{j}}\,\, \prod_{i =m}^{N- 1} \, \, \xi_{i}^{\frac{\delta_{i} -1}{2(\delta_{i} +1)}}\, \, \overline{\nu}_{max}, \qquad\qquad m = 2,\, 3,\, .\,.\,.\, N-2,\, N-1.
\label{num}
\end{equation}
The different frequencies are illustrated in Fig. \ref{FIG1} with the dashed lines 
are therefore in the following hierarchy: 
\begin{equation}
\nu_{max} = \nu_{1} > \nu_{2} > \nu_{3} > \,.\,.\,.\,.> \nu_{N-2} > \nu_{N-1} > \nu_{N} = \nu_{r}.
\label{nununu}
\end{equation}
Th result of Eq. (\ref{nununu}) is a direct consequence of the monotonic 
shape of $a\, H$ for $a> a_{1}$. If the profile of $a\, H$ is not 
monotonic for $a> a_{1}$, the hierarchy between the different 
frequencies of the spectrum is different as it happens when 
there is a second inflationary stage of expansion between $a_{1}$ and 
$a_{r}$ \cite{ST4}.

\subsection{Local maxima of the spectral energy density}
Since the typical frequencies probed by space-borne interferometers are below the Hz 
the most interesting situation, from the practical viewpoint, involves a maximum of $h_{0}^2 \Omega_{gw}(\nu,\tau_{0})$  just before the final dominance of radiation when $a_{N}= a_{r}$.
In this case the extremum of the spectral energy density occurs for $\nu_{N-1}$.
Further maxima can also arise for higher frequencies and are more constrained by the measurements of the pulsar timing arrays and by the limits coming from wide-band detectors. 
Recalling the notations of Fig. \ref{FIG1} together with the explicit expressions of the slopes given in Eqs. (\ref{THREE2}) and (\ref{THREE2a}), the spectral energy density has a maximum for $\nu = \nu_{N-1}$ provided 
\begin{equation}
\delta_{N-2} \geq 1, \qquad\qquad \delta_{N-1} < 1 \qquad \Rightarrow \qquad n_{T}^{(N-2)} \leq 0 , \qquad\qquad n_{T}^{(N-1)} > 0,
\label{nunu2}
\end{equation}
where $\nu_{N-2} > \nu_{N-1} > \nu_{N}= \nu_{r}$.
The spectral energy density in critical units reaches therefore a maximum for $\nu = \nu_{N-1}$ and its value is:
\begin{equation}
h_{0}^2\,\Omega_{gw} = \overline{{\mathcal N}}_{\rho}(\nu_{N-1}, r_{T}) \biggl(\frac{\nu_{N-1}}{\nu_{N}}\biggr)^{n_{T}^{(N-1)}}.
\label{nunu3}
\end{equation}
where $ \overline{{\mathcal N}}_{\rho}(\nu, r_{T})$ is a function that weakly depends on the frequency and it is 
typically smaller than $10^{-16}$ for $r_{T} \leq 0.06$ \cite{TS1,TS2,TS3}. 
This function is explicitly determined in section \ref{sec5} and it contains 
the dependence upon the transfer functions of the problem. 
According to the results deduced so far the explicit form of $\nu_{N-1}$ and $\nu_{N}$ is given by:
\begin{eqnarray}
\nu_{N-1} &=& \sqrt{\xi_{1}}\,.\,\,.\,\,.\,\,\sqrt{\xi_{N-2}} \,\, \xi_{N-1}^{\frac{\delta_{N-1}-1}{2 (\delta_{N-1} +1)}} \,\, \overline{\nu}_{max},
\label{nunu4}\\
\nu_{N} &=& \nu_{r} = \sqrt{\xi_{1}}\,.\,\,.\,\,.\,\,\sqrt{\xi_{N-1}}\,\, \overline{\nu}_{max}.
\label{nunu5}
\end{eqnarray}
 The ratio of Eqs. (\ref{nunu4}) and (\ref{nunu5}) gives exactly the term appearing in Eq. (\ref{nunu3})  so that the value of the spectral energy density at the maximum can also be written as:
 \begin{equation}
h_{0}^2\,\Omega_{gw}(\nu_{N-1},\tau_{0}) \simeq \overline{{\mathcal N}}_{\rho}(\nu_{N-1}, r_{T}) \,\,\xi_{N-1}^{ - \frac{2 (1 -  \delta_{N-1})}{\delta_{N-1} +1}}.
\label{nunu6}
\end{equation}
The function 
$\overline{{\mathcal N}}_{\rho}(\nu_{N-1}, r_{T})$ is weakly dependent on the frequency and its explicit form is discussed in the following section. 
Since the local maximum for $\nu = \nu_{N-1}$ does not depend on different maxima possibly 
arising for $\nu < \nu_{N-1}$ the simplest situation, for the present purposes, is the one where $N=3$. In this case there are only two successive stages characterized by $ \delta_{1}$ and $\delta_{2}$. The maximal frequency of the spectrum is given by: 
\begin{equation}
\nu_{1} = \nu_{max} = \xi_{1}^{\frac{\delta_{1} -1}{2(\delta_{1} +1)}}\, \, 
 \xi_{2}^{\frac{\delta_{2} -1}{2(\delta_{2} +1)}}\, \, 
\overline{\nu}_{max}.
\end{equation}
The frequencies $\nu_{2}$ and $\nu_{3} = \nu_{r}$ are instead given by:
\begin{equation}
\nu_{2} = \sqrt{\xi_{1}} \, \, \xi_{2}^{\frac{\delta_{2} -1}{2(\delta_{2} +1)}} \,\,\overline{\nu}_{max}, \qquad\qquad \nu_{r} = \nu_{3} = \sqrt{\xi_{1}} \, \,\sqrt{\xi_{2}} \,\,\overline{\nu}_{max}.
\label{nu2nur}
\end{equation}
We have just have one peak for $\nu=\nu_{2}$ and Eq. (\ref{nunu6}) gives 
\begin{equation}
h_{0}^2\,\Omega_{gw}(\nu_{2},\tau_{0}) \simeq \,\overline{{\mathcal N}}_{\rho}(\nu_{2}, r_{T}) \,\,\xi_{2}^{ - \frac{2 (1 -  \delta_{2})}{\delta_{2} +1}}.
\label{nunu7}
\end{equation}
All in all, while the existence of an early stage of accelerated expansion 
is motivated by general requirements directly related to causality, 
the post-inflationary expansion history is not constrained 
prior to big-bang nucleosynthesis. The results obtained in this 
section are therefore applicable to any post-inflationary expansion 
rate and do not assume the dominance of radiation between $H_{1}$ and $H_{r}$.

\subsection{General requirements on the total number of $e$-folds}
As repeatedly stressed we always considered hereunder the possibility
that $H_{r} > 10^{-44} \, M_{P}$ suggesting that the plasma is already 
dominated by radiation for temperatures that are well above the MeV
as it happens, for instance, when the reheating stage is triggered 
by the decay of a gravitationally coupled massive scalar field. There are however some  possibilities 
where the MeV-scale reheating temperature could be induced by long-lived massive species with masses close to the weak scale, as suggested in Refs. 
 \cite{MEV1,MEV2}. In spite of this interesting option we simply regard the condition  $H_{r} \geq 10^{-44} \, M_{P}$ as an absolute lower limit on $H_{r}$. 
Indeed the gravitational waves only couple 
to the expansion rate and our purpose here is just to propose 
a framework where the early thermal history of the plasma could be tested 
via the spectra of the inflationary gravitons. 

Along this perspective it is useful to remark that the maximal number of inflationary $e$-folds accessible to large-scale observations can be different \cite{ST1} (see also \cite{ST3,ST4,HOR}) depending on the post-inflationary expansion history. The maximal number of $e$-folds  presently accessible to large-scale observation (${\mathcal N}_{max}$ in what follows) 
is computed by fitting the (redshifted) inflationary event horizon inside the current Hubble patch;
in other words we are led to require, in terms of Fig. \ref{FIG1}, that 
$H_{1}^{-1} (a_{0}/a_{1}) \simeq H_{0}^{-1}$. It is clear that ${\mathcal N}_{max}$ 
{\em does not} coincide with the {\em total} number of $e$-folds that can easily be
larger (or even much larger) than ${\mathcal N}_{max}$. Depending on the various $\delta_{i}$ and $\xi_{i}$ 
the same gap in $a\, H$ is covered by a different amount of redshift. In the general situation 
of Fig. \ref{FIG1} the expression of ${\mathcal N}_{max}$ is given by:
\begin{eqnarray}
{\mathcal N}_{max}  &=& 61.88 - \ln{\biggl(\frac{h_{0}}{0.7}\biggr)} + \frac{1}{4} \ln{\biggl(\frac{r_{T}}{0.06}\biggr)} +\frac{1}{4} \ln{\biggl(\frac{{\mathcal A}_{{\mathcal R}}}{0.06}\biggr)}
\nonumber\\
&+&   \frac{1}{4} \ln{\biggl(\frac{h_{0}^2 \, \Omega_{R0}}{0.06}\biggr)}
+ \frac{1}{2}\sum_{i}^{N-1} \, \biggl(\frac{\delta_{i} -1}{\delta_{i} + 1}\biggr) \, \ln{\xi_{i}}.
\label{Nmax}
\end{eqnarray}
In connection with Eq. (\ref{Nmax}) we have three complementary possibilities.
If we conventionally set $\delta_{i} =1$ into Eq. (\ref{Nmax}) we obtain the 
standard result  implying that ${\mathcal N}_{max} = {\mathcal O}(60)$. 
Recalling that all the $\xi_{i}$ are, by definition, all smaller than $1$ we have that ${\mathcal N}_{max} > 60$ if $\delta_{i}<1$. For the same reason ${\mathcal N}_{max} < 60$ iff $\delta_{i}>1$. Let us consider, for instance, the case of a single phase expanding slower than radiation; in this case ${\mathcal N}_{max}$ can be as large as $75$ \cite{ST3,ST3a}. In the intermediate situations where there are different phases expanding either faster or slower than radiation ${\mathcal N}_{max}$ depends on the relative duration of the various phases and on their expansion rates. 

\renewcommand{\theequation}{4.\arabic{equation}}
\setcounter{equation}{0}
\section{The frequency range of space-borne interferometers}
\label{sec4}
\subsection{Approximate frequencies of the various instruments}
While various  space-borne interferometers have been proposed so far the presumed sensitivity of these instruments is still under debate. For this reason 
we adopt here a pragmatic viewpoint based on the considerations developed after Eq. (\ref{ONEeq}). In short the strategy is the following:
\begin{itemize}
\item{} the fiducial frequency interval of space-borne interferometers ranges from a fraction 
 of the mHz to the Hz and, within this interval,  the minimal detectable spectral energy density (denoted hereunder by $h_{0}^2 \Omega_{gw}^{(min)}(\nu, \tau_{0})$) defines 
 the potential sensitivity of the hypothetical instrument;
 \item{}  the LISA interferometers \cite{LISA1,LISA2} might hopefully probe the following region 
 of the parameter space:
\begin{equation}
h_{0}^2 \Omega_{gw}^{(min)}(\nu, \tau_{0}) = {\mathcal O}(10^{-11.2}), \qquad \qquad 10^{-4} \mathrm{Hz} < \nu \leq 0.1 \, \mathrm{Hz};
\label{SB1}
\end{equation}
\item{} in the case of the Deci-Hertz Interferometer Gravitational Wave Observatory (DECIGO) \cite{DECIGO1,DECIGO2} the minimal 
detectable spectral energy density could be smaller 
\begin{equation}
10^{-17.5} \leq h_{0}^2 \Omega_{gw}^{(min)}(\nu, \tau_{0}) \leq {\mathcal O}(10^{-13.1}), \qquad \qquad 10^{-3} \mathrm{Hz} < \nu \leq 0.1 \, \mathrm{Hz}.
\label{SB2}
\end{equation}
\end{itemize}
The values of Eq. (\ref{SB2}) are still quite hypothetical so that it is prudent to choose $h_{0}^2 \Omega_{gw}^{(min)}(\nu, \tau_{0})$
between the standard values of the hoped sensitivity of the DECIGO project 
 \cite{DECIGO1,DECIGO2} (suggesting $h_{0}^2 \Omega_{gw}^{(min)}(\nu, \tau_{0}) = {\mathcal O}(10^{-13.1})$) 
 and the optimistic figure reachable by the Ultimate-DECIGO \cite{UDECIGO} (conventionally 
 referred to as U-DECIGO) where  $h_{0}^2 \Omega_{gw}^{(min)}(\nu, \tau_{0}) = {\mathcal O}(10^{-17.5})$. For the record, the Big Bang Observer (BBO) \cite{BBO} might reach sensitivities 
 \begin{equation}
 h_{0}^2 \Omega_{gw}^{(min)}(\nu, \tau_{0})=  {\mathcal O}(10^{-14.2}), \qquad  10^{-3} \mathrm{Hz} < \nu \leq 0.1 \, \mathrm{Hz}.
\label{SB3}
\end{equation}
There finally exist also recent proposals such as Taiji \cite{TAIJI1,TAIJI2} and  TianQin \cite{TIANQIN1,TIANQIN2}
leading to figures that are roughly comparable with the LISA values. In summary for the typical frequency of the space-borne detectors we 
consider the following broad range:
\begin{equation}
0.1\, \mathrm{mHz} <\nu_{S} < 0.1 \, \mathrm{Hz}
\label{FOUR1}
\end{equation}
and suppose that in the range (\ref{FOUR1})
$h_{0}^2 \Omega_{gw}^{(min)}(\nu, \tau_{0})$  may take the following two extreme 
values.
\begin{equation}
h_{0}^2 \Omega_{gw}^{(min)}(\nu_{S}, \tau_{0}) = {\mathcal O}(10^{-11}), \qquad \qquad h_{0}^2 \Omega_{gw}^{(min)}(\nu_{S}, \tau_{0}) = {\mathcal O}(10^{-14}).
\label{SB4}
\end{equation}
While the two values of Eq. (\ref{SB4}) are both quite optimistic,  they are customarily assumed by the observational proposals and, for this reason, they are used here only for illustration.

\subsection{The profile of the spectral energy density}
The exclusion plots 
 characterizing the parameter space of the model are separately 
considered for the two illustrative values of $h_{0}^2 \Omega_{gw}^{(min)}(\nu_{S},\tau_{0})$ given in Eq. (\ref{SB4}).
For instance in Fig. \ref{FIG2} we require that 
\begin{equation}
\nu_{N-1} = {\mathcal O}(\nu_{S}), \qquad\qquad h_{0}^2 \Omega_{gw}(\nu_{N-1}, \tau_{0}) \geq 10^{-11}.
\label{FOUR1a}
\end{equation}
The first requirement of Eq. (\ref{FOUR1a}) implies that the frequency range of the maximum is comparable 
with $\nu_{S}$ while the second condition just comes from Eq. (\ref{SB4}) and it also demands, incidentally,  that the inflationary signal is larger than the 
spectral energy density produced by the gravitational waves associated with a 
putative strongly first-order phase transition, as we shall briefly discuss later on. The condition (\ref{FOUR1a})
can also be relaxed by assuming the second value of $h_{0}^{2} \Omega_{gw}(\nu_{S}, \tau_{0})$:
\begin{equation}
\nu_{N-1} = {\mathcal O}(\nu_{S}), \qquad\qquad h_{0}^2 \Omega_{gw}(\nu_{N-1}, \tau_{0}) \geq 10^{-14}.
\label{FOUR1b}
\end{equation}
Equation (\ref{FOUR1b}) is justified by the nominal sensitivity of other space-borne interferometers such as DECIGO \cite{DECIGO1,DECIGO2} or U-DECIGO \cite{UDECIGO}.  To investigate the phenomenological implications 
 the simplest choice is to posit $N= 3$. In this case we just have one maximum for $\nu_{r}<\nu < \nu_{max}$  and the discussion of the parameters is therefore simpler even if, as already mentioned,
the essential features remain the same also in more complicated situations.
For $N=3$ the spectral energy density of the model is:
\begin{equation}
h_{0}^2 \, \Omega_{gw}(\nu, \tau_{0}) = {\mathcal N}_{\rho} \, r_{T}(\nu_{p}) \biggl(\frac{\nu}{\nu_{p}}\biggr)^{\overline{n}_{T}(r_{T})} \,\, {\mathcal T}^2_{low}(\nu/\nu_{eq}) \,\,{\mathcal T}^2_{high}(\nu, \nu_{2}, \nu_{r}, n_{T}^{(1)}, n_{T}^{(2)}),
\label{FOUR2}
\end{equation}
where $\overline{n}_{T}(r_{T})$ has been computed in Eqs. (\ref{THREE2aa}) and (\ref{conf}); $\overline{n}_{T}$ is the spectral index associated with the wavelengths leaving the Hubble radius during the inflationary phase and reentering during the radiation stage. In Eq. (\ref{FOUR2}) $\nu_{p}$ and $\nu_{eq}$ define the lowest frequency range of the spectral energy density:
\begin{eqnarray}
\nu_{p} &=& \frac{k_{p}}{2\pi} = 3.092 \biggl(\frac{k_{p}}{0.002 \,\, \mathrm{Mpc}^{-1}}\biggr) \, \mathrm{aHz},
\nonumber\\
\nu_{eq} &=&  \frac{k_{\mathrm{eq}}}{2 \pi} = 15.97 \biggl(\frac{h_{0}^2\,\Omega_{M0}}{0.1411}\biggr) \biggl(\frac{h_{0}^2\,\Omega_{R0}}{4.15 \times 10^{-5}}\biggr)^{-1/2}\,\, \mathrm{aHz},
\label{FOUR4}
\end{eqnarray}
where we used that $k_{eq} = 0.0732\, h_{0}^2\,\Omega_{M0}\, \mathrm{Mpc}^{-1}$ (as usual, $\Omega_{M0}$ is the present fraction in dusty matter). 
The spectral slopes $n_{T}^{(1)}$ and $n_{T}^{(2)}$ are instead determined 
by Eqs. (\ref{THREE2}) and (\ref{THREE2a});  up to corrections ${\mathcal O}(r_{T})$ we have 
\begin{equation}
n_{T}^{(1)} = 2 ( 1 - \delta_{1}) + {\mathcal O}(r_{T}), \qquad\qquad 
n_{T}^{(2)} = 2 ( 1 - \delta_{2}) + {\mathcal O}(r_{T}),
\end{equation}
where $n_{T}^{(1)} < 0$ and $n_{T}^{(2)} >0$ since 
during the first stage the Universe expands faster than radiation 
(i.e. $\delta_{1}>1$) while in the second stage it is 
slower than radiation (i.e. $\delta_{2}<1$). In the simplest 
case where the consistency relations are enforced we have that
\begin{equation}
\overline{n}_{T}(r_{T}) = - \frac{r_{T}}{8} + {\mathcal O}(r_{T}^2), \qquad\qquad {\mathcal N}_{\rho} = 4.165 \times 10^{-15} \biggl(\frac{h_{0}^2\,\Omega_{R0}}{4.15\times 10^{-5}}\biggr).
\label{FOUR3}
\end{equation}
In Eq. (\ref{FOUR4}) ${\mathcal T}_{low}(\nu/\nu_{eq})$ is the low-frequency 
transfer function of the spectral energy density \cite{LIGO3}:
\begin{equation}
{\mathcal T}_{low}(\nu, \nu_{eq}) = \sqrt{1 + c_{1}\biggl(\frac{\nu_{eq}}{\nu}\biggr) + c_{2}\biggl(\frac{\nu_{eq}}{\nu}\biggr)^2},\qquad c_{1}= 0.5238, \qquad c_{2}=0.3537.
\label{FOUR5}
\end{equation}
The high-frequency transfer function ${\mathcal T}_{high}(\nu, \nu_{2}, \nu_{r}, \delta_{1}, \delta_{2})$ appearing 
in Eq. (\ref{FOUR2}) depends on $\nu_{2}$ and $\nu_{r}$ and it is given by:
\begin{equation}
{\mathcal T}_{high}^2(\nu,\nu_{r},\nu_{2},n_{T}^{(1)}, n_{T}^{(2)})) = \frac{\sqrt{1 + b_{1} (\nu/\nu_{r})^{n_{T}^{(2)}} + b_{2} (\nu/\nu_{r})^{2 n_{T}^{(2)}} }}{\sqrt{1 + d_{1} (\nu/\nu_{2})^{n_{T}^{(2)} +|n_{T}^{(1)}|} + d_{2} (\nu/\nu_{2})^{2(n_{T}^{(2)} +|n_{T}^{(1)}|)} }},
\label{FOUR6}
\end{equation}
where $b_{i}$ and $d_{i}$ (with $i = 1,\, 2$) are numerical coefficients of order $1$ that depend on the specific choice of $\delta_{1}$ and $\delta_{2}$ and cannot be written in general terms. We recall that the explicit expressions of $\nu_{2}$ and $\nu_{r}$ 
have been given in Eq. (\ref{nu2nur}) and they depend explicitly upon $\xi_{1}$ and $\xi_{2}$.
Since  Eq. (\ref{FOUR6}) depends on {\em two 
different scales},  there are {\em three relevant limits} of ${\mathcal T}_{high}^2(\nu,\nu_{r},\nu_{2})$ that must be considered. The first limit stipulates that:
\begin{equation}
{\mathcal T}_{high}^2(\nu,\nu_{r},\nu_{2}, n_{T}^{(1)}, n_{T}^{(2)}) \to \sqrt{\frac{b_{2}}{d_{2}}}\,\,\biggl(\frac{\nu_{2}}{\nu_{r}}\biggr)^{n_{T}^{(2)}} \,\,\biggl(\frac{\nu}{\nu_{2}}\biggr)^{- |n_{T}^{(1)}|}, \qquad\qquad \nu \gg \nu_{2},
\label{firstlim1}
\end{equation}
and it corresponds to the high-frequency branch where 
the spectral energy density is suppressed as $\nu^{-|n_{T}^{(1)}|}$:
\begin{equation}
h_{0}^2\,\Omega(\nu,\tau_{0}) = \overline{{\mathcal N}}_{\rho}(r_{T},\nu) \biggl(\frac{\nu_{2}}{\nu_{r}}\biggr)^{n_{T}^{(2)}}\biggl(\frac{\nu}{\nu_{2}}\biggr)^{-|n_{T}^{(1)}|}, \qquad\qquad \nu_{2} < \nu < \nu_{max}.
\label{firstlim2}
\end{equation}
Note that since $n_{T}^{(1)}<0$, in the spectral energy density we 
introduced the absolute value just to avoid potential confusions. 
This is not necessary in the case of $n_{T}^{(2)}$ which is instead positive semidefinite.
From Eq. (\ref{FOUR6}) the second relevant limit corresponds to the 
region where $\nu < \nu_{2}$:
\begin{equation}
{\mathcal T}_{high}^2(\nu,\nu_{r},\nu_{2}, n_{T}^{(1)}, n_{T}^{(2)}) \to \sqrt{b_{2}} \biggl(\frac{\nu}{\nu_{r}}\biggr)^{n_{T}^{(2)}},\qquad\qquad \nu_{r}< \nu< \nu_{2}.
\label{secondlim1}
\end{equation}
In this case the spectral energy density increases as $\nu^{n_{T}^{(2)}}$ and its 
approximate expression is given by:
\begin{equation}
h_{0}^2\,\Omega(\nu,\tau_{0}) = \overline{{\mathcal N}}_{\rho}(r_{T},\nu) \biggl(\frac{\nu}{\nu_{r}}\biggr)^{n_{T}^{(2)}}, \qquad\qquad \nu_{r} < \nu < \nu_{2}.
\label{FOUR8}
\end{equation}
The third relevant limit of the transfer function is finally for $\nu< \nu_{r}$
and, in this limit, Eq. (\ref{FOUR6}) simply goes to $1$:
\begin{equation}
{\mathcal T}_{high}^2(\nu,\nu_{r},\nu_{2}, n_{T}^{(1)}, n_{T}^{(2)}) \to  1, \qquad\qquad \nu< \nu_{r}.
\label{FOUR7}
\end{equation}
The function $\overline{N}_{\rho}(\nu, r_{T})$ has been already introduced in Eq. (\ref{nunu3}) and, as anticipated,  its explicit expression depends in fact 
upon the low-frequency transfer function:
\begin{equation}
 \overline{{\mathcal N}}_{\rho}(r_{T}, \nu) = {\mathcal N}_{\rho} \, r_{T} \biggl(\frac{\nu}{\nu_{p}}\biggr)^{\overline{n}_{T}} \, \, {\mathcal T}^2_{low}(\nu_{r}/\nu_{eq}), \qquad \qquad \frac{ d \ln{\overline{{\mathcal N}}_{\rho}}}{ d \ln{\nu}} = 
 - \frac{r_{T}}{8} \ll 1.
 \label{FOUR9}
 \end{equation}
 Even though the prefactor $\overline{{\mathcal N}}_{\rho}(r_{T}, \nu)$ has a mild frequency dependence coming from neutrino free-streaming, for simplified analytic estimates this dependence can be ignored, at least approximately; this in fact 
 the meaning of the second relation in Eq. (\ref{FOUR9}).
Along this perspective we can estimate $\overline{{\mathcal N}}_{\rho} = {\mathcal O}(10^{-16.5})$ for $r_{T} =0.06$. 

\subsection{The constrained parameter space}
\begin{figure}[!ht]
\centering
\includegraphics[height=7.3cm]{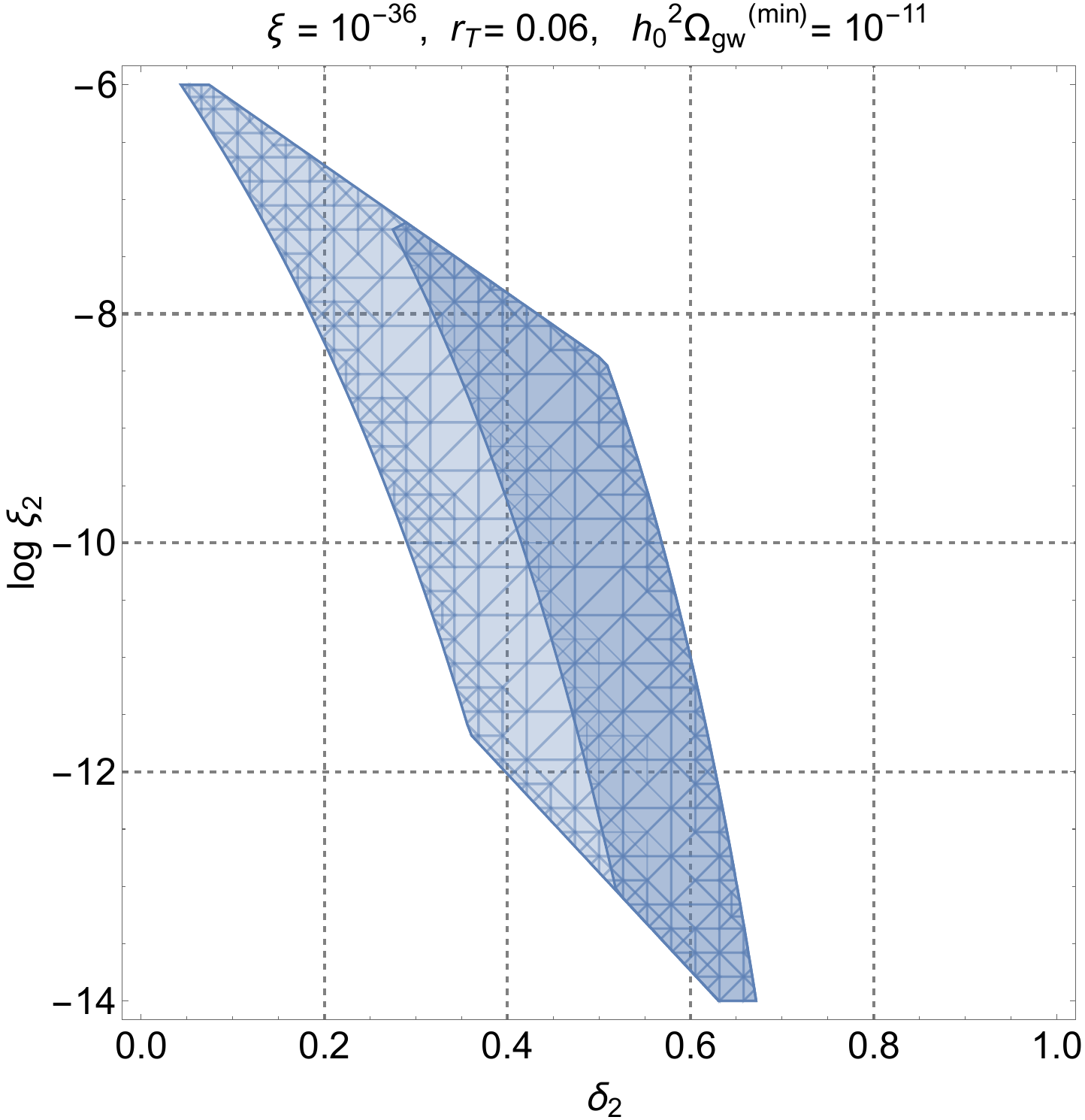}
\includegraphics[height=7.3cm]{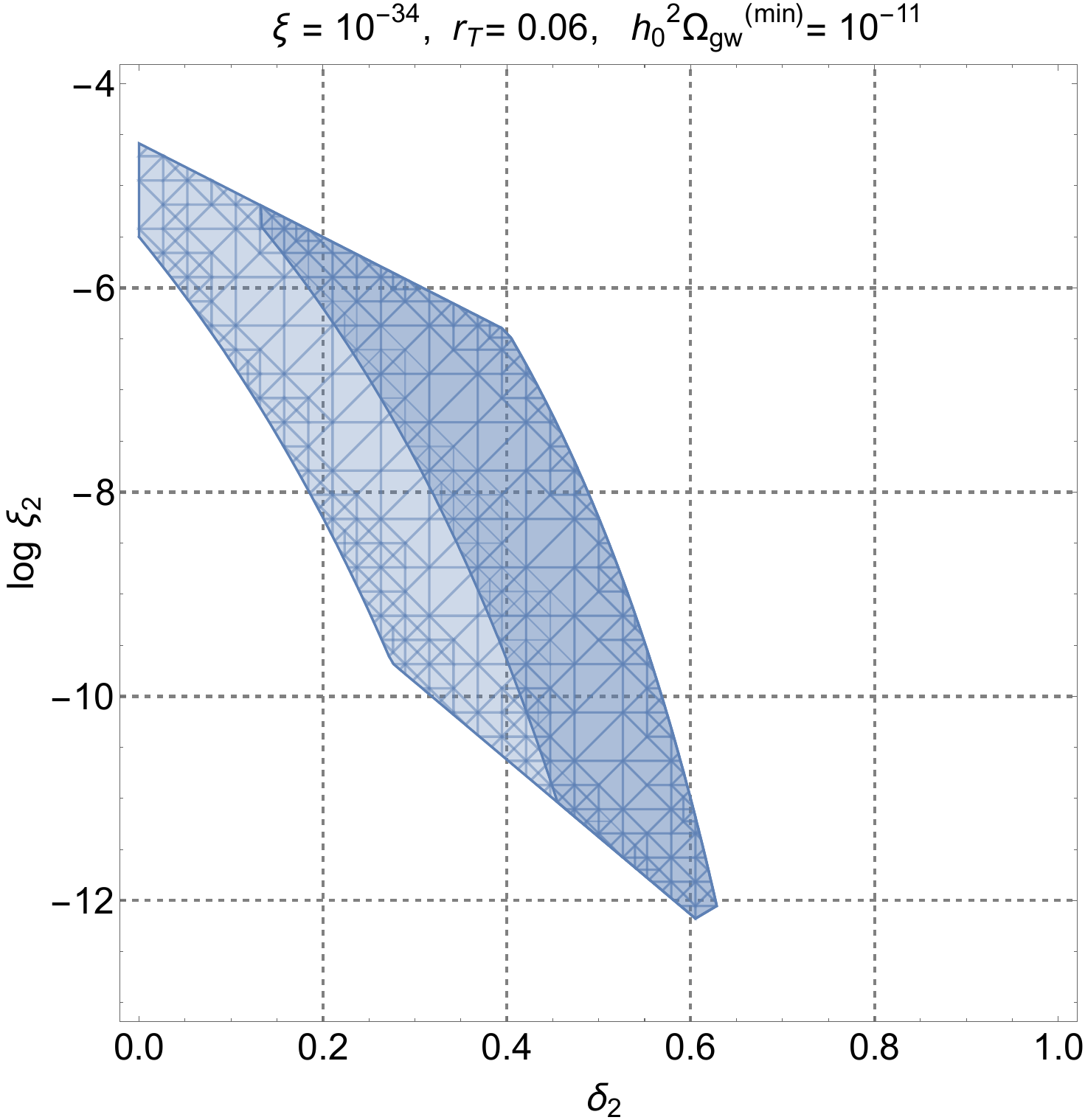}
\includegraphics[height=7.3cm]{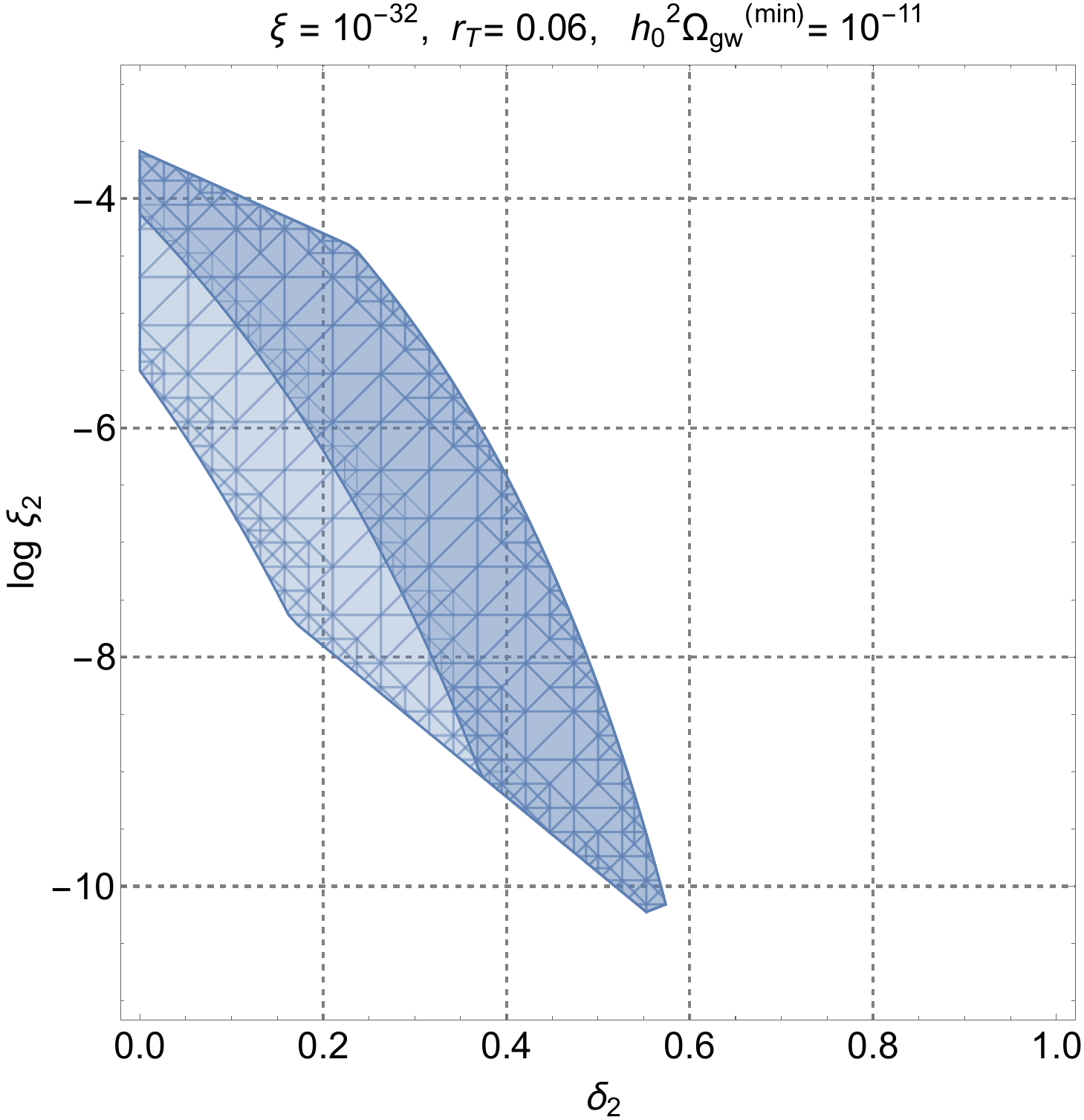}
\includegraphics[height=7.3cm]{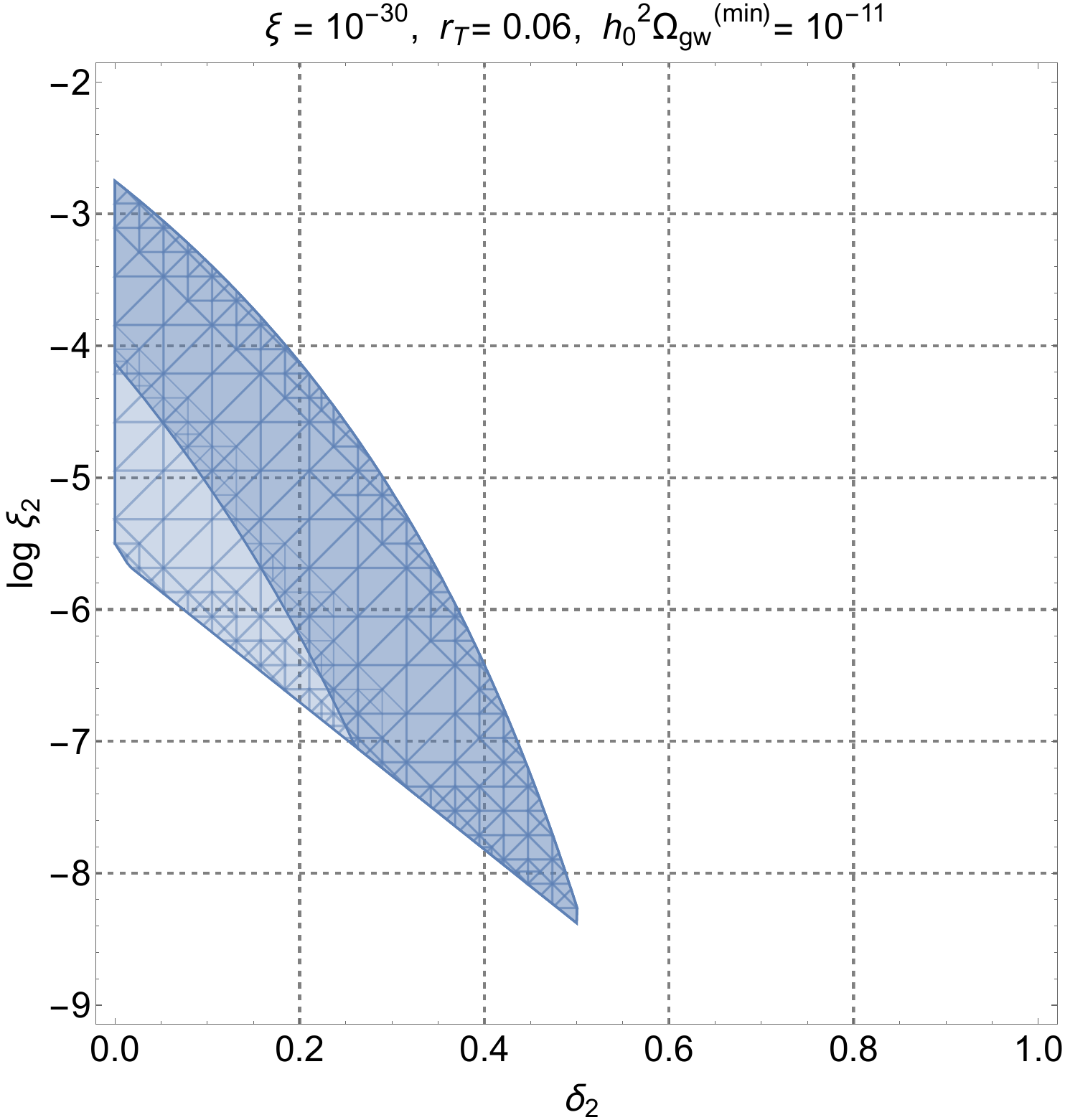}
\caption[a]{The parameter space is illustrated in the plane 
$(\delta_{2}, \, \xi_{2})$ by fixing $\delta_{1}$ to its most constraining 
value realized in the case $\delta_{1}\to 1$. In the four different plots of this figure the value of $\xi$ progressively 
increases between $10^{-38}$ and $10^{-30}$. The shaded area corresponds 
to the allow region. The wider area is obtained by enforcing the big-bang nucleosynthesis 
limit of Eqs. (\ref{BB1})--(\ref{BB2}) while the smaller (and darker) region follows by imposing the limits coming from the audio band (see Eqs. (\ref{CONS2})) and discussion therein.}
\label{FIG2}      
\end{figure}

The shaded region n Fig. \ref{FIG2} illustrates the area of the parameter space where the following pair of conditions are simultaneously verified:
\begin{equation}
h_{0}^2 \Omega_{gw}(\nu_{2}, \tau_{0}) \geq h_{0}^2 \Omega_{gw}^{(min)}, \qquad \qquad 
0.1\,\mathrm{mHz} < \nu_{2} \leq 0.1 \mathrm{Hz}.
\label{COND1}
\end{equation}
Since  the product of the various $\xi_{i}$ from $1$ to $(N-1)$ 
must equal $\xi$, in the case of $N = 3$ we have that $ \xi_{1} \xi_{2} = \xi$. 
Then the analysis can be simplified by using three related observations:
\begin{itemize}
\item{} $\xi_{1}$ can be traded for $\xi/\xi_{2}$ by recalling that, to avoid problems with nucleosynthesis,  the lower limit $\xi >10^{-38}$ must always be separately imposed;
\item{} as a consequence  we end-up with three parameters $\xi_{2}$, $\xi$, $\delta_{1}$ and 
$\delta_{2}$;
\item{} to discuss the parameter space we can fix $\delta_{1}\to 1$: this 
is the most constraining value since for $\delta_{1}> 1$ the high-frequency 
part of $h_{0}^2\,\Omega_{gw}(\nu,\tau_{0})$ decreases and it is therefore 
less constrained by the high-frequency limits.
\end{itemize}
If the spectral energy density decreases for $\nu> \nu_{2}$ (i.e. 
$n_{T}^{(1)} < 0$) all the high-frequency bounds (and in particular 
the LIGO-Virgo-KAGRA limit \cite{LIGO1,LIGO2,LIGO3}) are automatically satisfied {\em provided} they are satisfied in the case $\delta_{1} \to 1$.
The value of $\delta_{1}$ affects the value of the high-frequency 
slope of the spectral energy density since, in this case, $n_{T}^{(1)}(r_{T},\delta_{1}) 
= {\mathcal O}(r_{T})$. 

\begin{figure}[!ht]
\centering
\includegraphics[height=7.3cm]{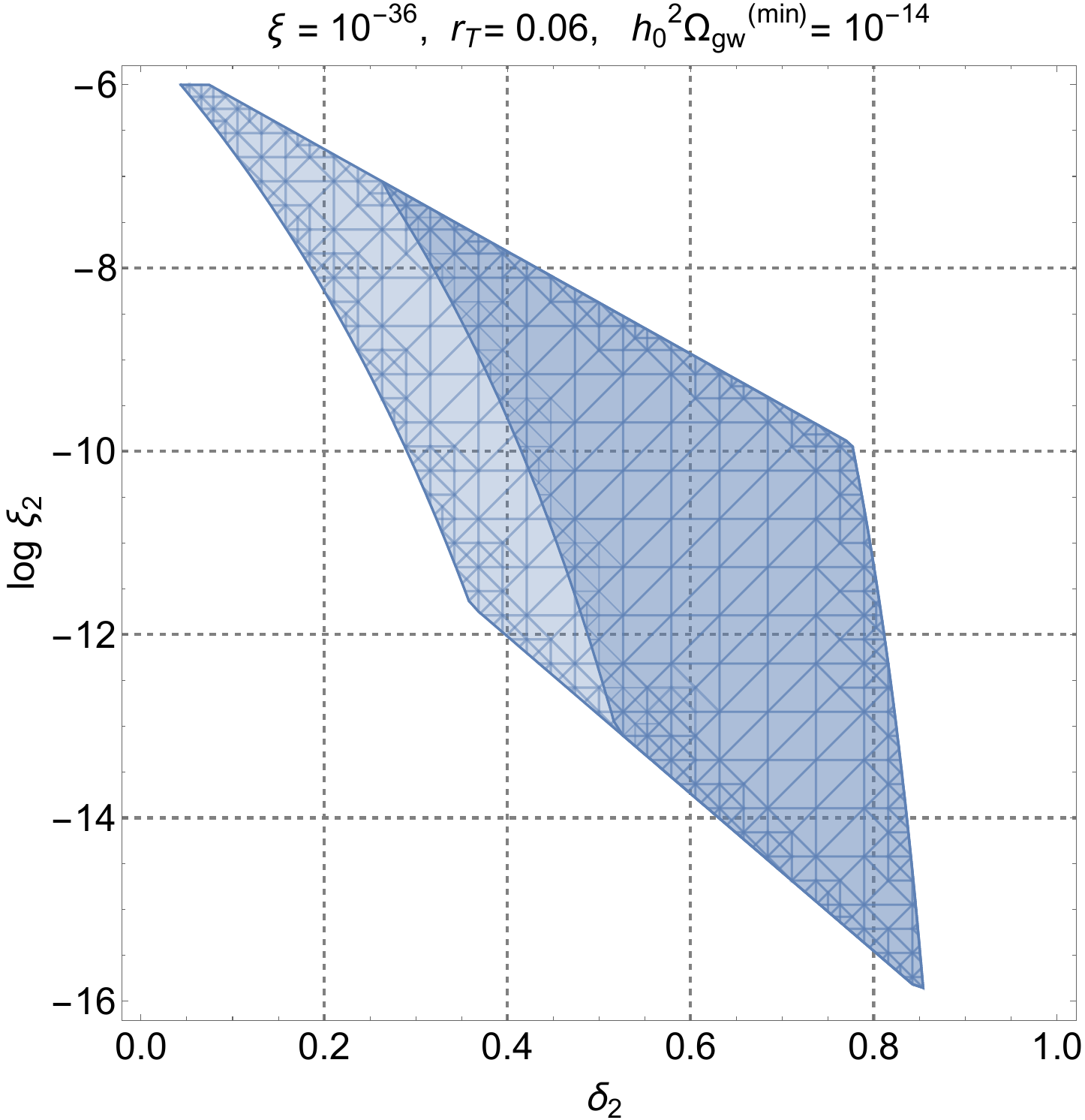}
\includegraphics[height=7.3cm]{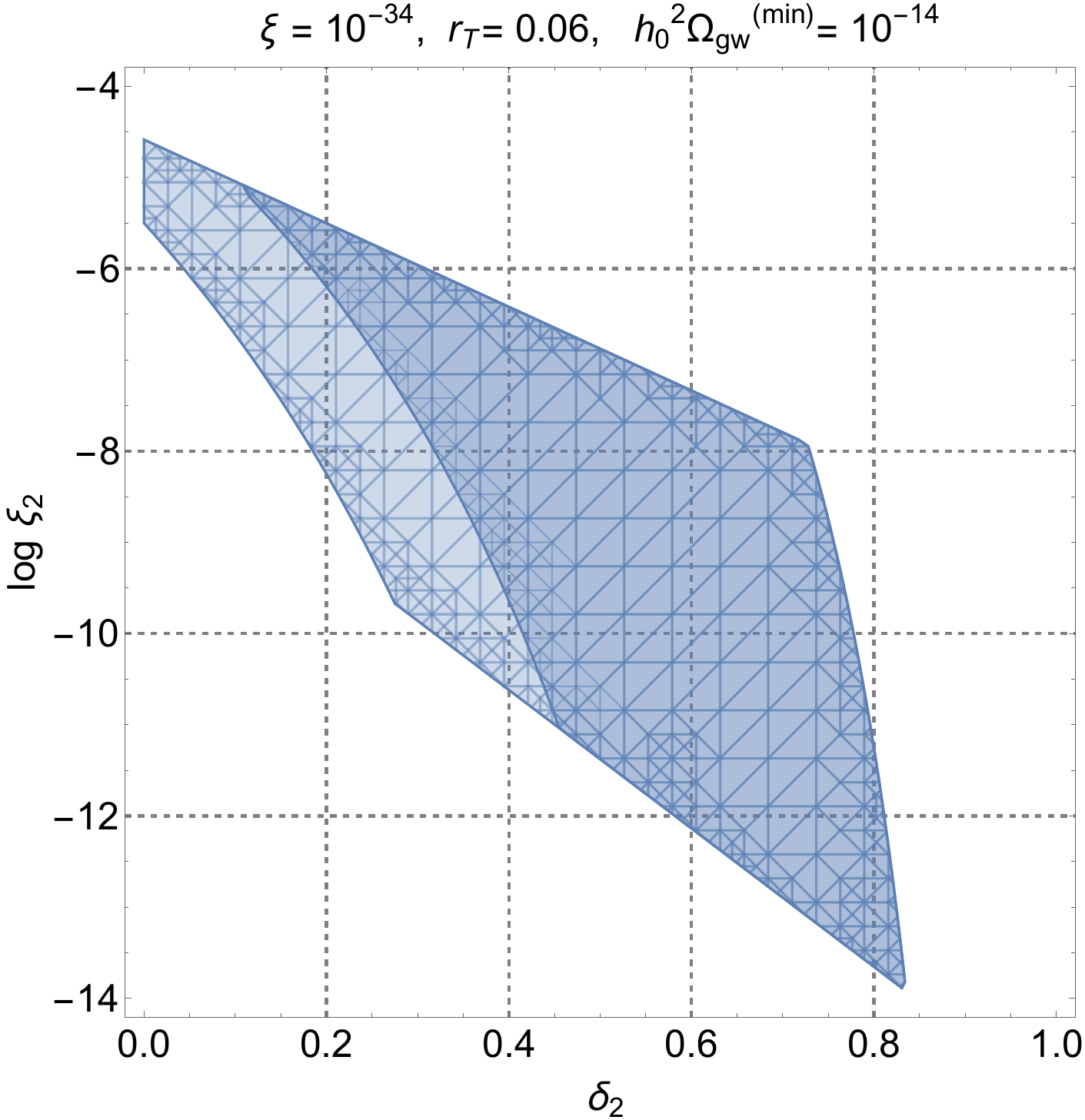}
\includegraphics[height=7.3cm]{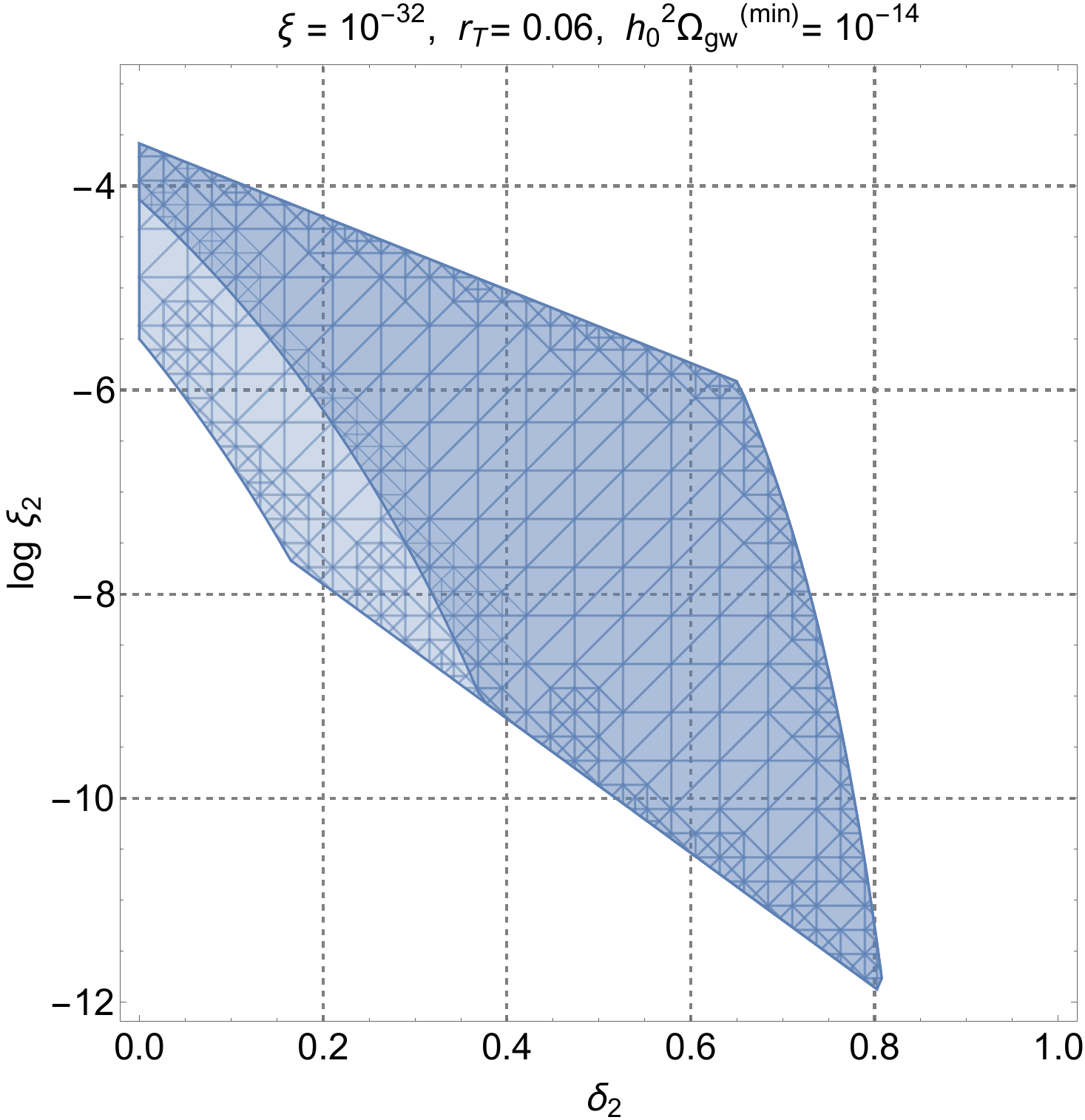}
\includegraphics[height=7.3cm]{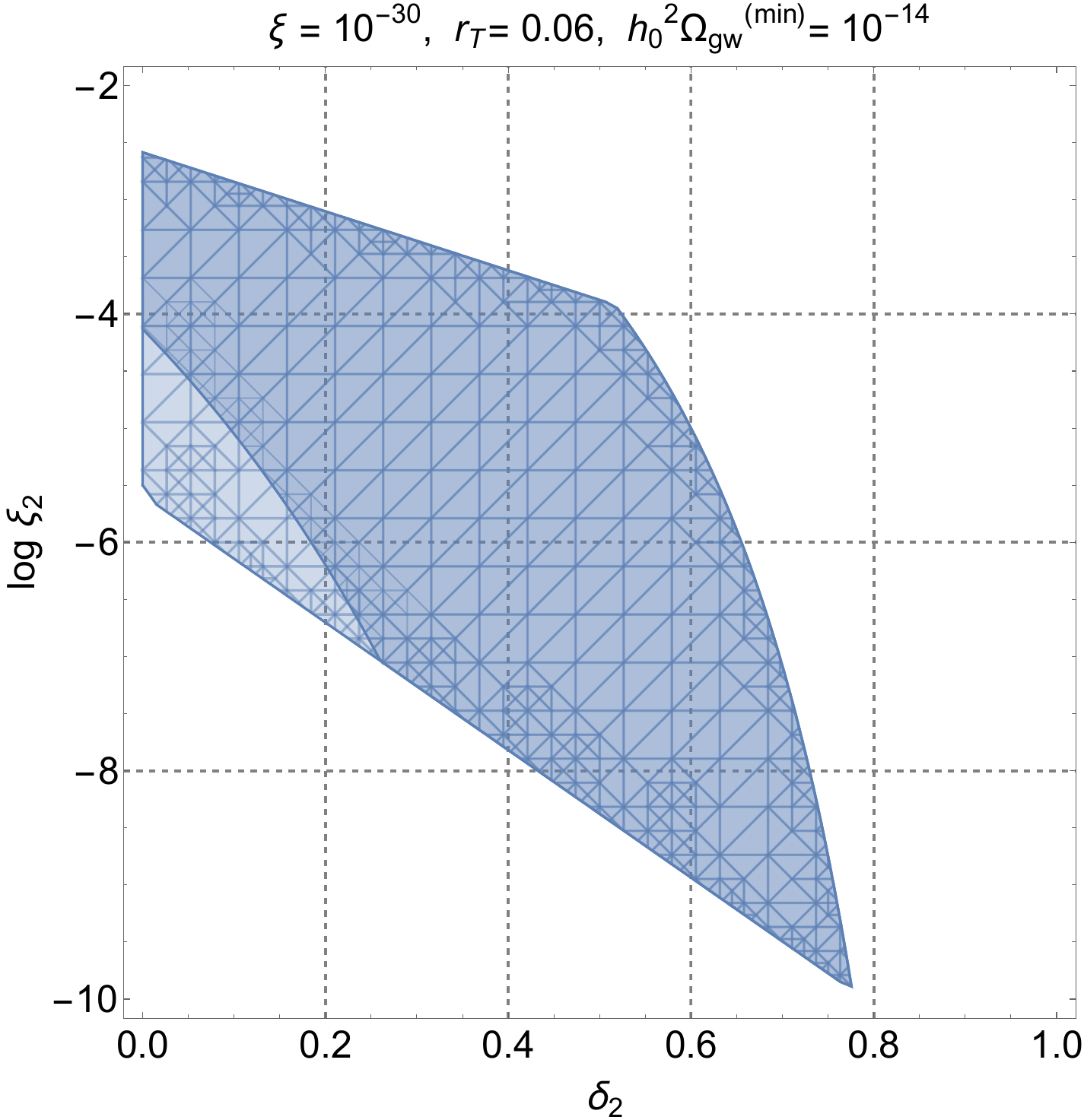}
\caption[a]{As in Fig. \ref{FIG2} the parameter space of the model is illustrated in the plane 
$(\delta_{2}, \, \xi_{2})$. The difference between these plots and the ones 
of Fig. \ref{FIG2} is related to $h_{0}^2 \Omega_{gw}^{(min)}$ which is now given by $10^{-14}$ while it was $10^{-11}$ in Fig. \ref{FIG2}.}
\label{FIG3}      
\end{figure}

In the four different plots of Fig. \ref{FIG2} the value of $\xi$ 
increases from $10^{-36}$ to $10^{-30}$ so that the radiation dominates 
when the expansion rate gets progressively larger; recall, in fact, that $\xi = H_{r}/H_{1}\geq 10^{-38}$. As $\xi$ increases the duration of the radiation phase increases, the shaded are gets smaller and 
allowed region is reduced.  The same logic of Fig. \ref{FIG2} 
has been followed in the case of Fig. \ref{FIG3}
with the difference that $h_{0}^2\Omega_{gw}^{(min)}$ is now relaxed from $10^{-11}$ to $10^{-14}$.  In each of the plots appearing in Figs. \ref{FIG2} and \ref{FIG3} there are two shaded regions. The wider 
area is obtained by enforcing the big-bang nucleosynthesis (BBN) constraint \cite{bbn1,bbn2,bbn3}.  The narrower (and darker) region in each plot of Figs. \ref{FIG2} and \ref{FIG3} is instead 
obtained by imposing the limits obtained from the operating interferometers on the 
backgrounds of relic gravitons, i.e. the LIGO-Virgo-KAGRA bound \cite{LIGO1} (see also \cite{LIGO2,LIGO3}). 
From the technical viewpoint 
the BBN constraint requires\footnote{The limit  of Eq. (\ref{BB2}) sets a constraint  on the extra-relativistic species possibly present at the BBN time. The limit is often expressed for practical reasons  in terms of $\Delta N_{\nu}$ representing the contribution of supplementary neutrino species. The actual bounds on $\Delta N_{\nu}$ range from $\Delta N_{\nu} \leq 0.2$ 
to $\Delta N_{\nu} \leq 1$;  the integrated spectral density in Eq. (\ref{BB2}) is thus between $10^{-6}$ and $10^{-5}$.}: 
\begin{equation}
h_{0}^2  \int_{\nu_{bbn}}^{\nu_{max}}
  \Omega_{gw}(\nu,\tau_{0}) d\ln{\nu} = 5.61 \times 10^{-6} \Delta N_{\nu} 
  \biggl(\frac{h_{0}^2\,\Omega_{\gamma0}}{2.47 \times 10^{-5}}\biggr),
\label{BB1}
\end{equation}
where $\Omega_{\gamma0}$ is the (present) critical fraction of CMB photons and $\nu_{bbn}$ denotes the BBN 
frequency:
\begin{equation}
\nu_{bbn}= 2.252\times 10^{-11} \biggl(\frac{N_{eff}}{10.75}\biggr)^{1/4} \biggl(\frac{T_{bbn}}{\,\,\mathrm{MeV}}\biggr) 
\biggl(\frac{h_{0}^2\,\Omega_{R0}}{4.15 \times 10^{-5}}\biggr)^{1/4}\,\,\mathrm{Hz},
\label{BB2}
\end{equation}
where  $N_{eff}$ denotes the effective number of relativistic degrees of freedom entering the total energy density of the plasma and $T_{bbn}$ is the temperature of big-bang nucleosynthesis. The bound (\ref{BB2}) can be relaxed 
 if the nucleosynthesis takes place in the presence 
of matter-antimatter domains \cite{bbn2} and $\nu_{max}$ appearing in Eq. (\ref{BB1}) 
denotes, as previously discussed, the maximal frequency of the spectrum. However, since 
$h_{0}^2 \Omega_{gw}(\nu, \tau_{0})$ decreases for $\nu > \nu_{2}$ (and more generally 
for $\nu > \nu_{N-1}$) the region between $\nu_{2}$ and $\nu_{max}$ gives a subleading contribution to the integral appearing at the left-hand side of Eq. (\ref{BB1}).
According to Figs. \ref{FIG2} and \ref{FIG3} the limits imposed by Eqs. (\ref{BB1})-- (\ref{BB2}) are less constraining than the ones following from the LIGO-Virgo-KAGRA bound\cite{LIGO1}. Indeed  the LIGO-Virgo-KAGRA collaboration, in its attempt to constrain the stochastic backgrounds of relic gravitons, 
 reported a constraint \cite{LIGO1} implying, in the case of a quasi-flat spectral energy density in the audio-band
\begin{equation}
\Omega_{gw}(\nu, \tau_{0}) < 5.8 \times 10^{-9}, \qquad\qquad 20 \,\, \mathrm{Hz} < \nu_{KLV} < 76.6 \,\, \mathrm{Hz},
\label{CONS2}
\end{equation}
where $\nu_{LVK}$ denotes the LIGO-Virgo-KAGRA frequency. The exclusion plots of Figs. \ref{FIG2} are then confronted with the current phenomenological bounds in all the available ranges of frequency with the aim of constraining the rate and the duration post-inflationary expansion Universe. In particular, in the nHz region, the pulsar timing arrays (PTA) recently reported a potential signal that could be attributed to the relic gravitons \cite{CCPP1,CCPP2,CCPP3,NANO1}.  
The PTA recently reported evidence 
of a potential signal in the nHz band. Using the spectral energy density in critical units as a pivotal variable the features of this purported signal would imply, in the present notations, that:
\begin{equation}
q_{0}^2\times 10^{-8.86}  < \, h_{0}^2\,\Omega_{gw}(\nu,\tau_{0}) < \,q_{0}^2 \times10^{-9.88} , \qquad\qquad 3\,\mathrm{nHz} < \nu< 100 \, \mathrm{nHz}.
\label{PTAb1}
\end{equation}
In Eq. (\ref{PTAb1}) we introduced  the numerical factor $q_{0}$ that depends on the 
specific experimental determination. The Parkes Pulsar Timing Array collaboration  \cite{CCPP1} suggests $q_{0}= 2.2$. Similarly the International Pulsar Timing Array collaboration (IPTA in what follows) estimates $q_{0}= 2.8$ \cite{CCPP2} while the European Pulsar Timing Array collaboration (EPTA in what follows) \cite{CCPP3} gives $q_{0} = 2.95$ (see also \cite{EPTA1,EPTA2}).  The results  of PPTA, IPTA and EPTA seem, at the moment, to be broadly compatble with the NANOgrav 12.5 yrs data \cite{NANO1} (see also \cite{NANO2,NANO3}) implying $q_{0} =1.92$. 

It is relevant to point out that neither the observations of Refs. \cite{CCPP1,CCPP2,CCPP3} nor the ones of Ref. \cite{NANO1} can be interpreted yet as an evidence of relic gravitons. The property of a PTA is that the signal from relic gravitons will be correlated across the baselines, while that from the other noise will not. Since these correlation have not been observed so far, the interpretation suggested in Eq. (\ref{PTAb1}) is still preliminary, to say the least.  To be fair the pragmatic strategy followed here 
will be to interpret Eq. (\ref{PTAb1}) as an upper limit whenever the 
corresponding theoretical signal is too low in the nHz region. Conversely  
if $h_{0}^2\,\Omega_{gw}(\nu,\tau_{0})$ happens to be grossly compatible 
with the range of Eq. (\ref{PTAb1}) it will be interesting to see if the associated spectral 
energy density fits within the PTA window.
The average of the $q_{0}$ of the different experiments is given by $\overline{q}_{0} = 2.46$. If we use $\overline{q}_{0}$ into Eq. (\ref{PTAb1}) we get  to the requirement: 
\begin{equation}
 10^{-8.07}  < \, h_{0}^2\,\Omega_{gw}(\nu,\tau_{0}) < \, \times10^{-9.09} , \qquad\qquad 3\,\mathrm{nHz} < \nu< 100 \, \mathrm{nHz}.
\label{PTAb1new}
\end{equation}
The condition (\ref{PTAb1new}) is never verified for the parameter space 
illustrated in Figs. \ref{FIG2} and \ref{FIG3}. Thus, for the selection 
of parameters analyzed here, $h_{0}^2 \Omega_{gw}(\nu,\tau_{0})$ 
may lead to a relevant signal below the Hz but not in the nHz band. 

\subsection{The spectral energy density and its signatures}
\begin{figure}[!ht]
\centering
\includegraphics[height=8cm]{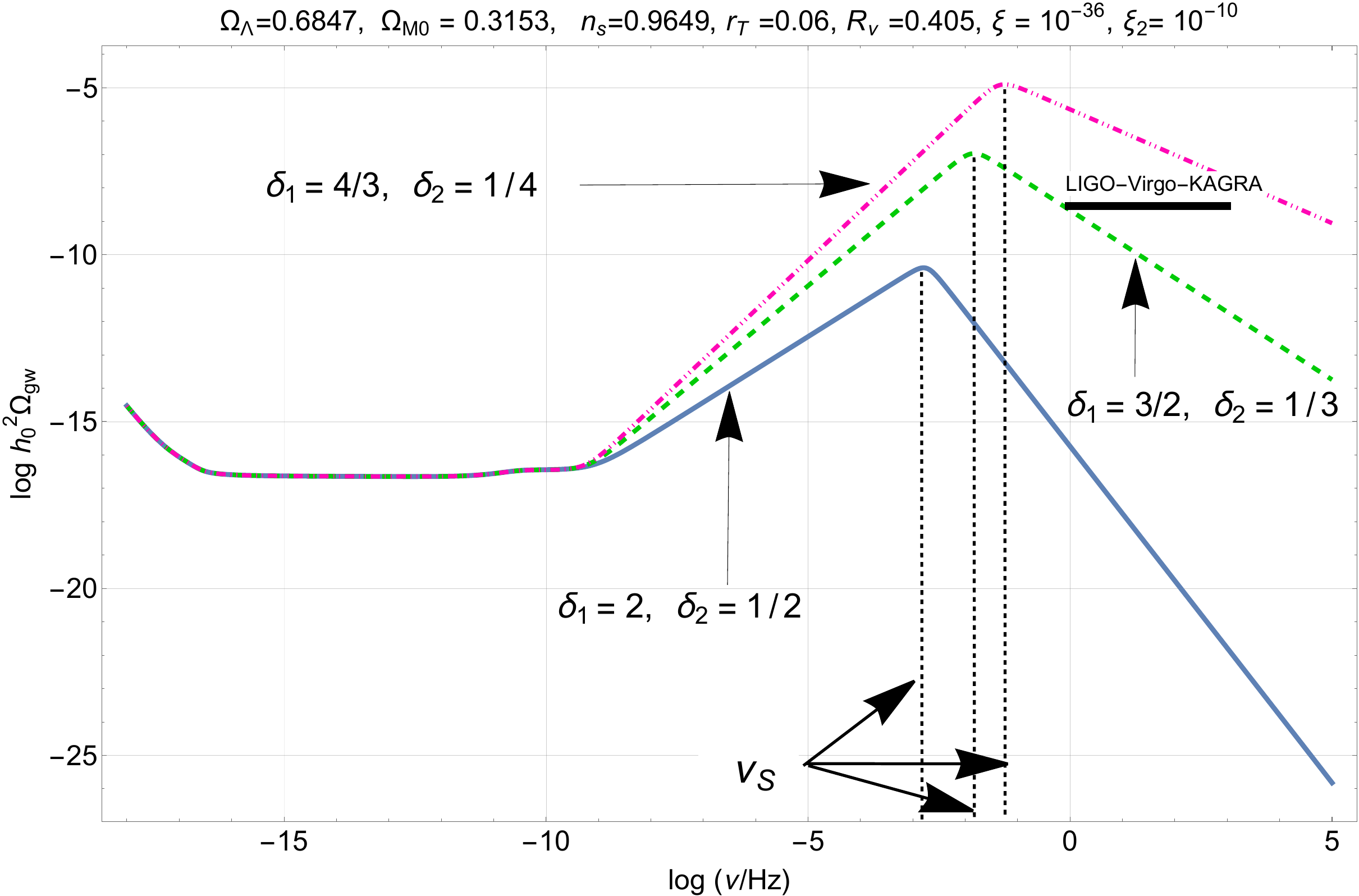}
\caption[a]{We illustrate the common logarithm of the spectral energy density of the 
relic gravitons as a function of the common logarithm of the frequency expressed in Hz. In this plot 
the dashed and the dot-dashed curves illustrate two models that are excluded by the big-bang 
nucleosynthesis constraint of Eq. (\ref{BB1}) and by the LIGO-Virgo-KAGRA limit of Eq. (\ref{CONS2}). The parameters of the curve at the bottom (full line) are instead drawn 
from the allowed region of the parameter space. In all the examples of this plot $\delta_{1}> 1$. }
\label{FIG4}      
\end{figure}

\begin{figure}[!ht]
\centering
\includegraphics[height=8cm]{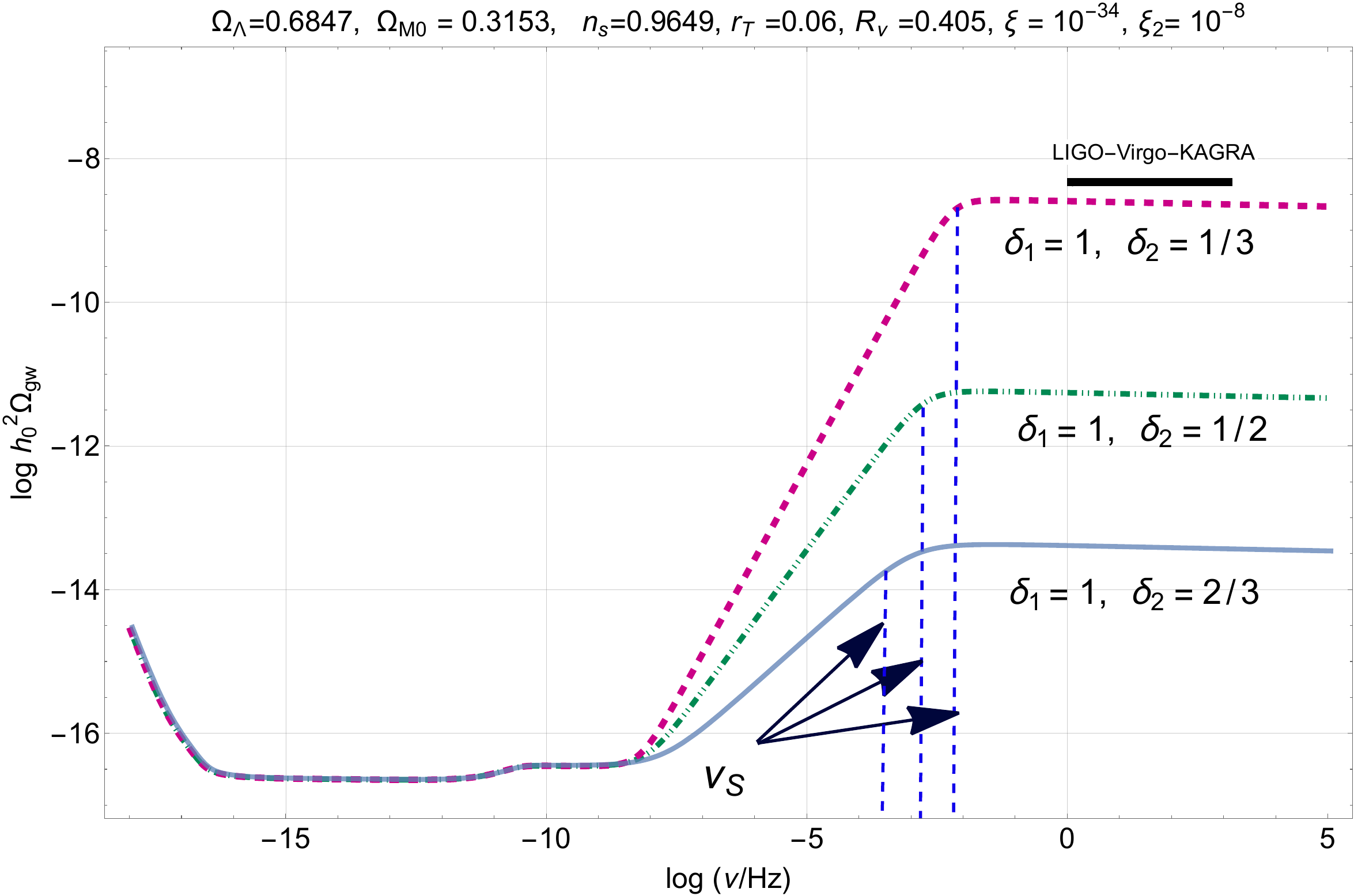}
\caption[a]{The conventions are exactly the same already explained in Fig. \ref{FIG4} but the spectral energy is illustrated for a different choice of the parameters. The dashed curves at the top of the figure is barely compatible with the limits set by Eq. (\ref{CONS2}). If $\delta_{1} > 1$ (see e.g. Fig. \ref{FIG4}) the spectral energy density decreases for $\nu> \nu_{2}$ and the most relevant constraint 
on the height of the maximum comes from Eqs. (\ref{BB1})--(\ref{BB2}). If 
$\delta_{1} \to 1$ the limits from the audio band (see, in particular, Eq. (\ref{CONS2})) 
are the most constraining ones and this 
is why we illustrated here this case where the spectral energy density remains quasi-flat at high-frequencies. }
\label{FIG5}      
\end{figure}

In Figs. \ref{FIG4} and \ref{FIG5}, the spectral energy density has been explicitly illustrated for a selection of the 
parameters.  
In Fig. \ref{FIG4} we selected $\xi = 10^{-36}$ and $\xi_{2} = 10^{-10}$ for different values of
$\delta_{1} > 1$ and $\delta_{2} < 1$. As expected the value of $\nu_{r}$ is always larger than $10^{-10}$. For this reason $h_{0}^2 \Omega_{gw}(\nu,\tau_{0})$ does not exceed ${\mathcal O}(10^{-16})$ in the frequency range of  the PTA and the profiles of  Figs. \ref{FIG4} and \ref{FIG5} are unable to account for the putative signal of Eqs. (\ref{PTAb1})--(\ref{PTAb1new}). Note, in this respect, that the parameters of the dot-dashed and of the dashed curves of Fig. \ref{FIG4} have been selected in order to get an artificially large signal that is in fact excluded both by the constraint of Eq. (\ref{BB1}) and by the limit of ground-based detectors (see Eq. (\ref{CONS2}) and discussion therein). Even in this case the PTA 
values are too large and must be accounted by different mechanisms.
The results of Fig. \ref{FIG5} correspond instead to a slightly different choice of the parameters, namely 
$\xi = 10^{-34}$ and $\xi_{2} =10^{-8}$. For illustration we have chosen $\delta_{1} \to 1$ implying 
that between $\nu_{max}$ and $\nu_{2}$ the spectral energy density is quasi-flat. This is the most
constraining case from the viewpoint of the limits coming from wide-band detectors \cite{LIGO1,LIGO2,LIGO3}.

The class of signals computed here are distinguishable, at least in principle, from the other 
astrophysical and cosmological foregrounds. In the region between a fraction of the mHz and the 
Hz the predominant astrophysical foregrounds of gravitational radiation 
are probably associated with the galactic distribution of the white dwarves. 
While the signal of white dwarves could also be used for calibration, other astrophysical 
foregrounds are also expected (e.g. stellar origin black holes and even supermassive black holes from galaxy mergers). 
Another cosmological foreground is given by TeV scale early Universe and this happens since the typical frequency corresponding to the Hubble radius 
at the electroweak stage is ${\mathcal O}(10) \, \mu\mathrm{Hz}$.  
To have drastic deviations from homogeneity and the consequent production of burst of gravitational radiation the electroweak phase transition must be strongly first-order. In this 
respect two classes of related observations are in order. 
The first remark is that the electroweak phase transition 
does not have to be strongly first-order. Since the first perturbative analyses of the problem we actually know that that for the measured range of Higgs masses the electroweak phase transition is not of first-order \cite{linde1,linde2}. Due to the inherently non-perturbative nature of the problem, the original perturbative estimates have been subsequently corroborated by lattice calculations both in three \cite{PT1} and four dimensions \cite{PT2}. 
For the measured values of the Higgs and gauge boson masses the transition between 
the symmetric and the broken phase follows a cross-over evolution that should 
not lead to an appreciable production of gravitational radiation. In this 
case the only cosmological signal may be the one associated 
with the inflationary gravitons.

The second remark is that, at the moment the hopes of observing a burst of gravitational radiation from the electroweak scale must rely on some extension of the standard electroweak theory. In this situation the amount of gravitational radiation 
produced by the phase transition depends on the particular model and also on the 
difference between the energy density in the broken and in the symmetric phase. This 
energy density may be comparable with the energy density of the ambient plasma 
(and in this case the phase transition experiences a strong supercooling) or smaller 
than the energy density of the surrounding radiation (and in this case the phase transition 
is mildly supercooled). If the gravitational is produced from the 
collisions of the bubbles of the new phase \cite{coll1,coll2,coll3} the equivalent $h_{0}^2 \Omega_{gw}(\nu,\tau_{0})$ scales like $\nu^3$, reaches a
maximum and then decreases with a power that may be faster than $\nu^{-1}$. The spectral energy density inherits also contribution from the sound waves of the plasma  
\cite{coll4} and this second component may be even larger than the one due to bubble collisions. The key point for the present ends is that the powers associated with a strongly first-order 
phase transition are typically much steeper than the ones discussed here.
In our case the rise of $h_{0}^2\Omega_{gw}(\nu,\tau_{0})$ appearing in Figs. \ref{FIG4} and \ref{FIG5} always scales with $0< n_{T}^{(2)}\leq 1$  while the corresponding slope in the case of phase transitions is typically ${\mathcal O}(3)$. Another possibility not requiring a strongly 
first-order phase transition is the presence of a stochastic 
background of hypermagnetic fields at the electroweak phase. In this case bursts 
of gravitational radiation may also be produced
and the spectral energy density is different from the one discussed here \cite{hyper1} (see also \cite{hyper2,hyper3}). Overall, because of causality, the spectra associated with the TeV physics are much steeper around their putative maximum. For this reason it seems plausible to disentangle the inflationary contributions from other possible cosmological foregrounds.
There is of course a deeper problem that has to do with our ability to 
separate the cosmological signals from the other astrophysical foregrounds
(e.g. white dwarves, massive and supermassive black-holes). 
The potential difficulties associated with the astrophysical foregrounds 
suggested already many years ago \cite{ST1,ST2} that the potential signals of post-inflationary stages should be probably observed over much 
higher frequencies ${\mathcal O}(\mathrm{MHz})$ where electromagnetic 
detectors might be operating in the future \cite{cav1,cav2,cav3,cav4,cav5,cav6}.

\subsection{Complementary considerations}
 Even if specific scenarios involving high, intermediate and low reheating temperatures have ben suggested in the past
\cite{ST1,ST2,PV} (see also, in this respect, Refs. \cite{ADD5,LL1,LL2,LL3,ford,spok,ST3,ST3a,ST4}),
the present analysis focussed on a model-independent perspective. 
Given that the early expansion history of the background is unknown the only plausible strategy is to combine 
the low-frequency limits and of the high-frequency constraints. This approach suggests that a signal below the
Hz  is not excluded and the features of the spectra can be clearly distinguished from the ones 
of strongly first-order phase transitions that are the main competitive signal in this region. We might think, by the same token, 
that large signals can be achieved also over much smaller frequencies and it is then interesting to apply the present model-independent strategy in this instance.

By looking at Figs. \ref{FIG4} and \ref{FIG5} we may note 
that the first break from scale invariance of the spectral energy density 
always occurs above a typical frequency ${\mathcal O}(10^{-10})$ Hz. This feature persists 
if the number of successive stages of expansion is increased and the ultimate reason for this 
occurrence is given by Eqs. (\ref{nur})--(\ref{nur2}): while all the intermediate frequencies 
of the spectrum are given by a complicated combination involving the various expansion rates 
and the intermediate curvature scales (see e.g. Eqs. (\ref{num})--(\ref{nununu}))
$\nu_{r}$ is solely determined by the ratio between $H_{r}$ and $H_{1}$.
This means that the absolute lower limit 
  $H_{r} \geq 10^{-44} \, M_{P}$ imposed by big-bang nucleosynthesis 
  also implies that $\nu_{r} > {\mathcal O}(10^{-10})$ Hz, or, more precisely $ \nu > \nu_{bbn}$.
This requirement is essential when combining the high-frequency limits on the relic gravitons 
with the low-frequency one \cite{ST3,ST3a} and it has been correctly implemented 
in a recent analysis focusing on the post-inflationary reheating parameters \cite{REQ1}.
A direct consequence of this requirement is that the recent results of the PTA (see Eq. (\ref{PTAb1new}))
cannot be explained by a post-inflationary modification of the expansion rate. Given the current 
limits on $r_{T}$ \cite{TS1,TS2,TS3} the largest 
value of $h_{0}^2 \Omega_{gw}$ for $\nu < \nu_{r}$ is, at most, ${\mathcal O}(10^{-16.5})$.
Since $\nu_{r} \geq {\mathcal O}(10^{-10})$ Hz it is impossible that $h_{0}^2 \Omega_{gw}$ 
reaches a value ${\mathcal O}(10^{-9})$ for typical frequencies of the order of $10$ nHz or even 
$100$ nHz as required for an explanation of the PTA observations \cite{CCPP1,CCPP2,CCPP3,NANO1}. We actually remind that the 
largest slope of the spectral energy density, in the case of a barotropic fluid, is of order $1$ and 
it occurs for a nearly stiff equation of state, as established long ago \cite{ST1,ST2}. 
There have been nonetheless claims of a sound explanation of the PTA data by post-inflationary 
stiff phases. For instance in Ref. \cite{REQ2} the authors just suggest the opposite of what we just 
said; by looking more carefully at the results\footnote{See, in particular, the two plots in Fig. 8 of Ref. \cite{REQ2}
where the lowest break of the spectrum is ${\mathcal O}(10^{-14})$ Hz and possibly even smaller.}
we see that, in this case, $\nu_{r} = {\mathcal O}(10^{-14})$ Hz. By appealing to the model-independent 
results of  Eqs. (\ref{nur})--(\ref{nur2}) that apply strictly also in this case we have
therefore 
\begin{equation}
\nu_{r} = {\mathcal O}(10^{-14}) \, \mathrm{Hz} \,\qquad \Rightarrow \qquad \xi = {\mathcal O}(10^{-45}),
\label{est1}
\end{equation}
and if we now recall that $\xi=H_{r}/H_{1}$ we the conclude that 
\begin{equation}
\frac{H_{r}}{M_{P}} = {\mathcal O}(10^{-51})  \biggl(\frac{r_{T}}{0.06}\biggr)^{1/2} \,\, 
\biggl(\frac{{\mathcal A}_{{\mathcal R}}}{2.41\times 10^{-9}}\biggr)^{1/2},
\label{est2}
\end{equation}
which is much smaller than the lower limit on $H_{r}$ imposed throughout this analysis. There 
is therefore no surprise that $h_{0}^2 \Omega_{gw}(\nu)$ could be as large as $10^{-10}$ in the 
nHz range since from $10^{-14}$ Hz to $10$ nHz there are $6$ orders of magnitude and 
$h_{0}^2 \Omega_{gw}(\nu)$ may increase, in this range, with linear slope from $10^{-16}$ to $10^{-10}$.
Reference \cite{REQ2} demands therefore that the plasma 
is {\em not} dominated by radiation by the time of big-bang nucleosynthesis and this 
approach is totally rejected by the viewpoint conveyed in the present and analysis\footnote{It should also 
be rejected on a more general ground since it is unclear how it is possible to form the light element abundances in such a context. We remark 
that the slopes derived by the authors in Eqs. (93)--(94) of Ref. \cite{REQ1} coincide exactly with the ones 
discussed long ago (see e.g. Eq. (3.32) of Ref. \cite{ST1}). }. 

The considerations developed in the previous paragraph are also essential for a fair comparison 
of the relic gravitons discussed here with the cosmic string signals (see, in this respect, the review of Ref. \cite{LIGO3}). The gravitational waves emitted by oscillating loops at different epochs have been argued to produce a stochastic background \cite{REQ3} with quasi-flat spectral energy density which is typically larger than the inflationary signal. The nature of the signal changes depending on three basic parameters: the string tension in Planck units (i.e. $G \mu$); the typical size of the loops normalized at the formation time; the emission efficiency of the loop.
The quoted values of $G\mu$ may range between $10^{-8}$ and $10^{-23}$ while the typical size of the loop may vary between $10^{-10}$ and $10^{-1}$. The large interval of variation of the parameters makes it obvious that different signals can be expected. From symmetry breaking in the grand unified context the typical values  of $G \mu$ could be as large as ${\mathcal O}(10^{-6})$. These values 
would cause however measurable temperature and polarization anisotropies of the CMB and have been ruled out; current limits from CMB observations demand $G\mu < {\mathcal O} (10^{-8})$. 
For the largest values of $G\mu$ potentially compatible with CMB data the $h_{0}^2 \Omega_{gw}$ from 
cosmic strings exhibits a hump in the nHz region \cite{REQ4} (see also \cite{REQ5}) and then flattens out.  As 
$G\, \mu$ diminishes the hump shifts at higher frequencies and the overall signal is suppressed 
potentially getting to $h_{0}^2 \Omega_{gw} = {\mathcal O}(10^{-15})$ for $G\,\mu = {\mathcal O}(10^{-21})$.
Since the relic gravitons discussed here never lead to a large signal in the nHz region, the only possible 
ambiguity may arise when $G \, \mu \ll 10^{-9}$. In this case the nearly flat branch 
of the cosmic string signal might be confused with situations similar to the one described, for instance, in Fig. \ref{FIG5}.
A detailed comparison is however beyond the scopes of this paper.

The final point we want to mention concerns the possibility of second-order effects and their interplay 
with the considerations presented here. In the concordance paradigm where 
the curvature inhomogeneities are Gaussian and adiabatic the  stochastic backgrounds of 
relic gravitons are corrected by second-order effects that involve an effective anisotropic stress \cite{REQ6}
which is however gauge-dependent\footnote{It has been noted that the different gauge-dependent 
results can be swiftly compared by a careful use of the normal modes of the system. It turns out that 
the results obtained in different gauges is comparable for typical wavelengths shorter than the Hubble radius. See, in this respect, the discussion in Refs. \cite{REQ7a,REQ7}.}.
The tensor modes reentering the Hubble radius when the plasma is dominated by a stiff fluid lead to a spectral energy density whose blue slope depends on the total post-inflationary sound speed. This result gets however corrected  by a secondary  
term coming from the curvature inhomogeneities that reenter all along the same stage of expansion. In comparison with the first-order result, the secondary contribution has been shown to be always suppressed inside the sound horizon  and its effect  on the total spectral energy density of the relic gravitons is therefore negligible for all phenomenological purposes \cite{REQ7}. The same conclusion applies also in the present situation. 

\renewcommand{\theequation}{5.\arabic{equation}}
\setcounter{equation}{0}
\section{Concluding remarks and future perspectives}
\label{sec5}
If  the wavelengths that left the Hubble radius 
during inflation reentered in the radiation-dominated stage of expansion 
the spectral energy density of the inflationary gravitons
is today quasi-flat for typical frequencies larger than $100$ aHz. 
Prior to nucleosynthesis the timeline of the expansion rate is however unknown 
and we considered here a post-inflationary evolution consisting of a sequence of stages expanding at rates that are alternatively faster and slower than radiation. As a consequence,
the spectral energy density can even be eight orders of magnitude larger than the conventional inflationary signal for frequencies between the $\mu$Hz and a fraction of the Hz. 

Below the Hz various space-borne detectors will probably be operational in the next twenty years and 
 the signals expected in the mHz region are dominated by astrophysical sources (e.g. 
galactic white dwarves, solar-mass black holes, supermassive black holes coming from galaxy mergers). The only cosmological sources considered in this context are associated with 
the phase transitions at the TeV scale even if it is well established, both 
perturbatively and non-perturbatively, that the standard electroweak theory 
leads to a cross-over regime where drastic deviations from homogeneity (and the 
consequent bursts of gravitational radiation) should not be expected. The inflationary signal is 
customarily regarded as irrelevant since its spectral energy density 
could be at most $h_{0}^2 \Omega_{gw}(\nu_{S}, \tau_{0})= {\mathcal O}(10^{-17})$ 
for $r_{T} \leq 0.06$ and for $\nu_{S} = {\mathcal O}(\mathrm{mHz})$. 
The relic gravitons of inflationary origin may instead lead to $h_{0}^2 \Omega_{gw}(\nu_{S}, \tau_{0})= {\mathcal O}(10^{-9})$ provided the expansion history prior 
to nucleosynthesis is not constantly dominated by radiation. 
The slopes of the spectral energy density obtained in the case of a putative strongly first-order phase transition are much steeper than the ones associated with a modified expansion 
history. When confronted with the most relevant phenomenological bounds
the class of signals discussed here is predominantly constrained by the limits on the massless species at the nucleosynthesis scale and by the direct observations of ground-based detectors (i.e. LIGO, Virgo and KAGRA). 
From the profiles of the spectral energy density and from the slopes of the hump in the mHz range it is possible to infer the post-inflationary expansion history for typical curvature scales that are between $10^{-44} \,\, M_{P}$ and $10^{-34} \,\, M_{P}$ (i.e. roughly $10$ orders of magnitude larger than the nucleosynthesis scale). The analysis of the spectral energy density in different frequency ranges (e.g. nHz, mHz and MHz) might even allow to reconstruct the expansion history of the Universe at earlier and later times. It is finally interesting that some regions of the parameter space that are relevant for space-borne detectors 
also lead to a potentially large signal in the audio band and will probably 
be directly probed or excluded in the near future.
 
All in all the perspective conveyed in this analysis suggests that the 
frequency range below the Hz should be carefully investigated in the light 
of a possible signal coming from inflationary gravitons. While a strongly 
first order phase transition may be realized beyond the standard electroweak 
theory, the present discussion only assumes a conventional inflationary 
stage supplemented by a post-inflationary evolution that  deviates from the 
conventional radiation dominance prior to  nucleosynthesis. The observations in the mHz region 
could then simultaneously test the occurrence of an early 
inflationary stage and of a post-inflationary expansion 
history whose details are still unknown and might only be discovered by looking 
at the spectra of relic gravitons.

\section*{Aknowledgements} 
It is a pleasure to thank T. Basaglia, A. Gentil-Beccot, S. Rohr, J. Vigen 
and the whole CERN Scientific Information Service for their kind help
during the preparation of this manuscript.

\end{document}